\documentclass{aastex62}
\usepackage{epsfig,amsmath,amsfonts,amssymb,graphicx,subfigure,enumitem,color}
\usepackage{amssymb}
\usepackage{amsmath}
\usepackage[varg]{txfonts}
\usepackage{natbib}
\usepackage{url}             
\usepackage{graphicx}
\usepackage{float}
\newcommand{\beq}{\begin{equation}}
\newcommand{\eeq}{\end{equation}}
\newcommand{\beqa}{\begin{eqnarray}}
\newcommand{\eeqa}{\end{eqnarray}}
\newcommand{\beqann}{\begin{eqnarray*}}
\newcommand{\eeqann}{\end{eqnarray*}}
\bibpunct{(}{)}{;}{a}{}{,} 

\usepackage{textcomp}
\newcommand{\foo}{\rlap{+}{\texttimes}}
\shorttitle{Rapid Forced Reconnection in Sun's Corona}
\shortauthors{Srivastava et al.}

\begin{document}
\title{On the observations of rapid forced reconnection in the solar corona}
\author{A.K.~Srivastava$^{1}$, S.K.~Mishra$^{1}$, P.~Jel\'inek$^{2}$, Tanmoy~Samanta$^{3}$, Hui~Tian$^{3}$, Vaibhav Pant$^{4}$, P.~Kayshap$^{2}$, Dipankar~Banerjee$^{5}$, J.G.~Doyle$^{6}$, B.N. Dwivedi$^{1}$}
\affil{$^{1}$Department of Physics, Indian Institute of Technology (BHU), Varanasi-221005, India.}
\affil{$^{2}$University of South Bohemia, Faculty of Science, Institute of Physics, Brani\v sovsk\'a 1760, CZ -- 370 05 \v{C}esk\'e Bud\v{e}jovice, Czech Republic.}
\affil{$^{3}$School of Earth and Space Sciences, Peking University, Beijing 100871, People's Republic of China.}
\affil{$^{4}$Centre for mathematical Plasma Astrophysics, Mathematics Department, KU Leuven, Celestijnenlaan 200B bus 2400, B-3001 Leuven, Belgium}
\affil{$^{5}$Indian Institute of Astrophysics, Kormangala, Bangalore, Karnataka, India.}
\affil{$^{6}$Armagh Observatory, College Hill, Armagh 9DG 73H, N. Ireland}
\bigskip
\begin{abstract}
Using multiwavelength imaging observations from the Atmospheric Imaging Assembly (AIA) onboard the Solar Dynamics Observatory (SDO) on 03 May 2012, we present a novel physical scenario for the formation of a temporary X-point in the solar corona, where plasma dynamics is forced externally by a moving prominence. Natural diffusion was not predominant, however, a prominence driven inflow occurred firstly, forming a thin current sheet and thereafter enabling a forced magnetic reconnection at a considerably high rate. Observations in relation to the numerical model reveal that forced reconnection may rapidly and efficiently occur at higher rates in the solar corona. This physical process may also heat the corona locally even without establishing a significant and self-consistent diffusion region. Using a parametric numerical study, we demonstrate that the implementation of the external driver increases the rate of the reconnection even when the resistivity required for creating normal diffusion region decreases at the X-point. We conjecture that the appropriate external forcing can bring the oppositely directed field lines into the temporarily created diffusion region firstly via the plasma inflows as seen in the observations. The reconnection and related plasma outflows may occur thereafter at considerably larger rates. 
\end{abstract}

\bigskip
\bigskip
\bigskip
\bigskip
\section{Introduction}
The million-degree hot solar corona maintains its high temperature and compensates for its radiative losses by continuously acquiring an energy flux of $\approx$10$^{3}$ W m$^{-2,}$. Recent studies suggest that energy transport in the solar corona is associated with localized magnetic flux-tubes, which can channel various kinds of magnetohydrodynamic (MHD) waves and shocks as heating candidates. Dissipation of electric current via magnetic reconnection  provides an alternate mechanism to heat the solar corona. However, there are various physical conditions that need to be established appropriately in the reconnection region to generate its high rate and subsequent energy release. 

As mentioned above, the Sun's corona continously requires a high amount of energy flux to compensate its radative losses (Withbroe \&  Noyes, 1977). The major energy sources in the localized solar corona may be associated with magneto-hydrodynamic (MHD) waves and shocks (e.g., De Pontieu, et al., 2007; Jess et al., 2009; McIntosh et al., 2011; Mart\'inez-Sykora et al., 2017; Srivastava et al., 2017,2018 and references cited therein). However, global magnetic reconnection is a well known mechanism to explain various physical processes in the universe, and laboratory plasma experiments, e.g., solar and stellar flares, geomagnetic substorms and magnetosphere of planets, tokamak disruptions, high energy fusion experiments, etc (e.g., Cargill, P.J. \& Klimchuk, 2004; Schwenn, 2006; Shibata \& Magara, 2011; Klimchuk, J.A., 2015, and references cited therein). In the astrophysical plasmas (e.g., solar corona), it is basically defined as a self-organization of the magnetic fields towards their more relaxed state and associated impulsive release of the stored magnetic energy (Yamada et al., 2010). In the solar corona, magnetic reconnection is considered as one of the major candidates to heat its atmosphere, and also to generate space weather candidates (e.g., flares and coronal mass ejections) that can affect the Earth's outer atmosphere, its satellite and communication system, etc (e.g., Cargill, P.J. \& Klimchuk, 2004; Schwenn, 2006; Shibata \& Magara, 2011; Klimchuk, J.A., 2015, and references cited therein). However, the main issues of magnetic reconnection are still unsettled despite several novel discoveries both in theory and observations in a variety of astrophysical and laboratory plasmas, e.g., formation of current sheet, appropriate reconnection rate, establishment of natural diffusion regions and their physical properties, etc (e.g., Yamada et al., 2010; Priest \& Forbes, 2007).
\\
\begin{figure*}
\begin{center}
\hspace{-0.8cm}
\includegraphics[scale=0.7,angle=0,width=17.cm,height=17.cm,keepaspectratio]{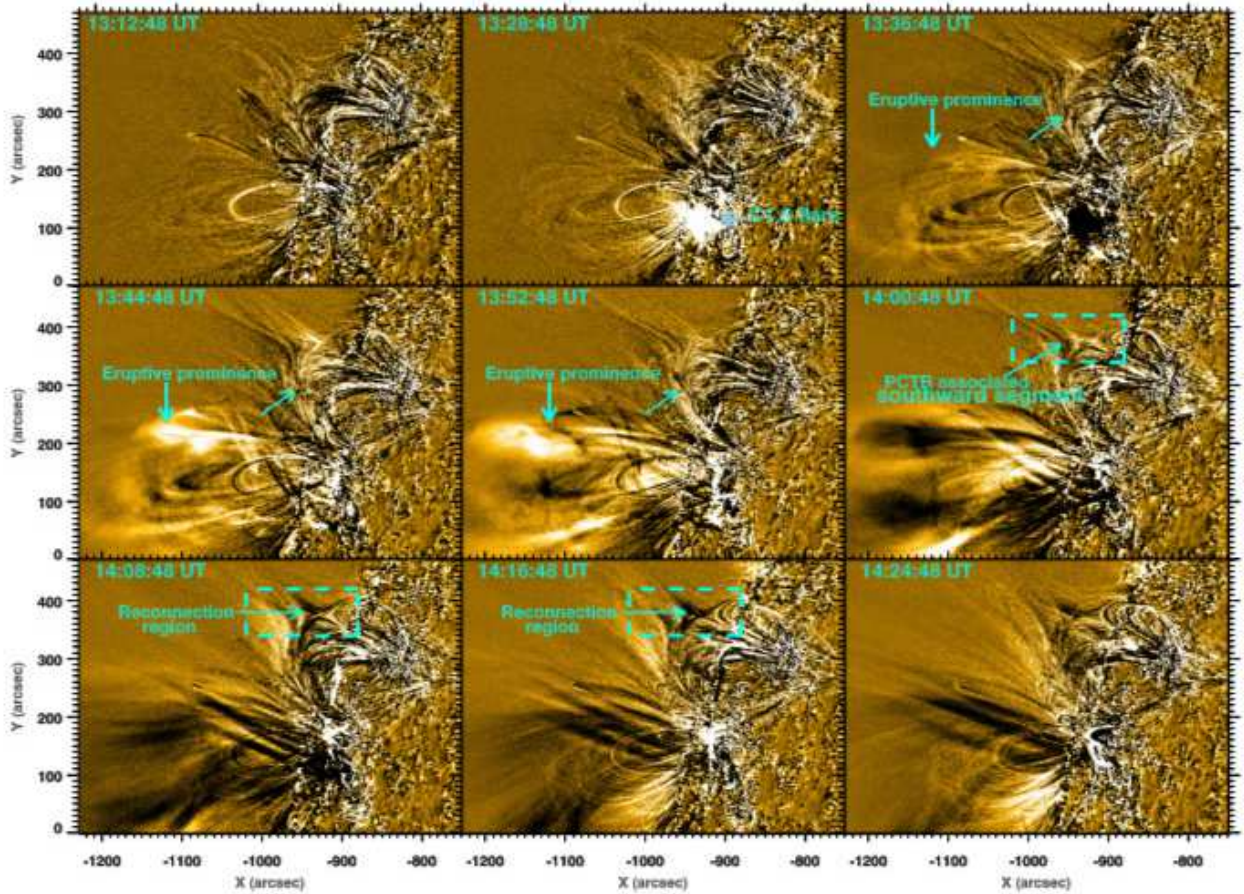}
\caption{The running difference image sequence in the SDO/AIA 171 $\AA$ shows an eruptive prominence and its impact on the surrounding coronal field lines. The eruptive prominence disturbs the surrounding magnetic fields (middle panel; indicated by cyan arrow), which is associated with a prominence. The prominence associated field lines are forced to bring the southward segment of prominence-corona transition region (PCTR) associated coronal field lines towards the reconnection-point. The reconnection region and region of interest (ROI) is indicated by  cyan box, where prominence driven southward segment of the field lines reconnect with the northward branch of the magnetic field anchored at the limb and form a temporary X-point. The online animation (fig1-anim.mp4) shows an eruptive prominence and its influence on surrounding coronal fields. This animation has a duration from 13:10 to 14:45 UT, and it shows a field-of-view containing a C-class flare, an associated prominence eruption, and the reconnection region.}
\end{center}
\end{figure*}

As mentioned in the above paragraph, the magnetic reconnection is introduced as breaking and reconfiguration of the oppositely directed magnetic field lines in highly conducting plasma. The magnetic field lines collapse near the X-point and form an extended magnetic singularities known as current sheet. There are two mechanism of the current sheet formation. The first kind of current sheet formation is associated with the MHD instabilities (e.g., resistive tearing mode and ideal kink mode) known as spontaneous magnetic reconnection (e.g., White 1984; Baty 2000; Vekstein 2017). The second kind of current sheet can be formed in the MHD stable configuration, where some external perturbations trigger the forced magnetic reconnection (Hahm \& Kulsrud 1984). The forced magnetic reconnection may be activated by nonlinear MHD waves, which may be caused by explosive solar activities (e.g., Sakai, Tajima \& Brunel 1984; Devar et al. 2013; Beidler et al. 2017). The forced magnetic reconnection may be developed due to boundary perturbations, which induce a surface current in such a way that it opposes the progress of the reconnection (Ishizawa \& Tokuda 2000, 2001; Fitzpatrick 2003). The multi-mode simulation approach has been adopted to investigate the thinning of the current sheet induced by forced magnetic reconnetion (Birn et al. 2005). The motion of the photospheric footpoints of the coronal magnetic field may also trigger the forced magnetic reconnection, which may be caused by the explosive solar coronal events (e.g., Vekstein \& Jain 1998; Jain et al. 2005; Vekstein 2017). Although, there is a remarkable development in the theory of the forced magnetic reconnection, Jess et al. (2010) have suggested that there is no observational evidence of explosive flare or coronal activities triggered by forced magnetic reconnection. They have observed a microflare activity driven by forced magnetic reconnection. The lower solar atmosphere (photosphere \& chromosphere) is dominated by cool, partially ionized and collision dominated plasma. The most of the energy release during the forced  magnetic reconnection may be consumed by such plasma systems (e.g., Litvinenko 1999; Chen et al. 2001; Chen \& Ding 2006; Litvinesko et al. 2007). \\
\begin{figure*}
\begin{center}
\hspace{-1cm}
\includegraphics[scale=0.6,angle=0,width=17cm,height=22.0cm,keepaspectratio]{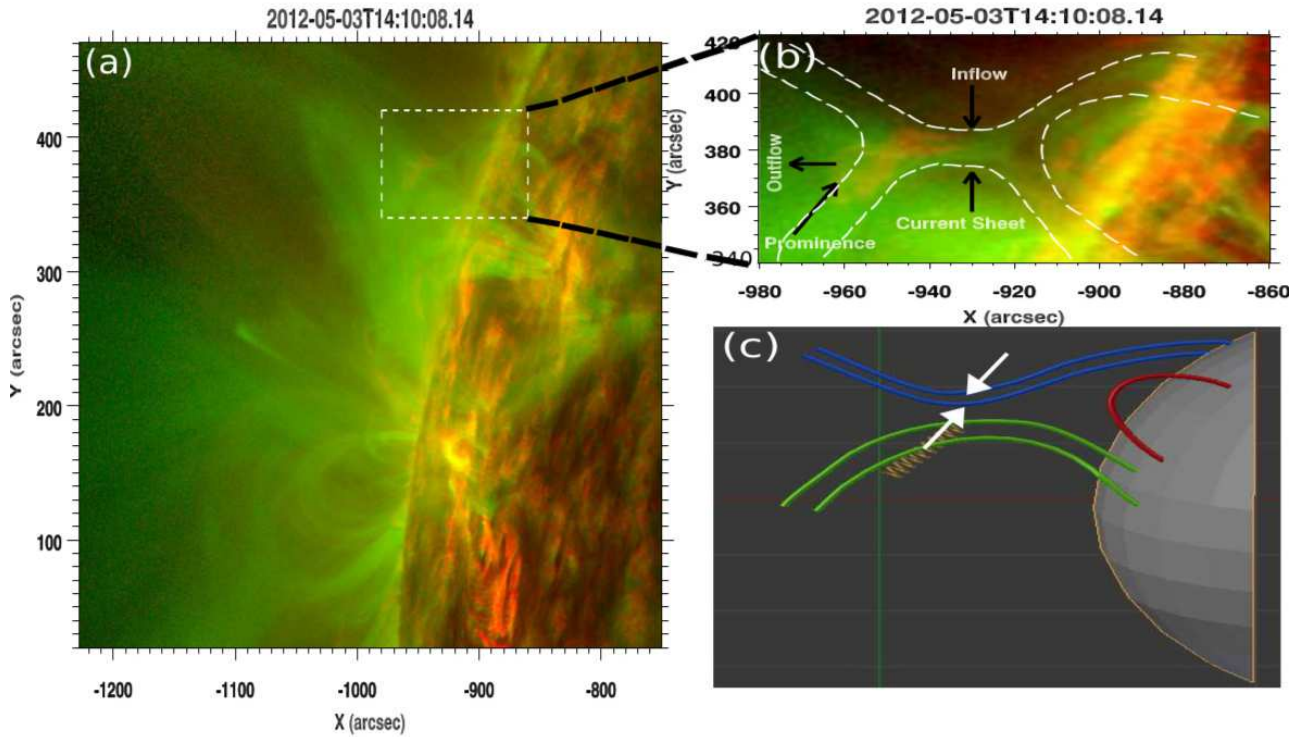}
\end{center}
\caption{\small \small Direct imaging of forced reconnection in the Sun's corona is demonstrated in this composite picture.
SDO/AIA imaging observations on 3$^{rd}$ May 2012 at 14:10:08 UT depict the formation of a temporary X-point and a forced reconnection region ($'$a$'$-$'$b$'$). The animations $'$fig2a-anim.mp4$'$ and $'$fig2b-anim.mp4$'$ show the triggering of the forced reconnection in the off-limb large-scale corona. They run during the time between 13:00 UT to 14:56 UT. The second animation shows the annotated prominence, inflows and outflows of the plasma, creation of an X-point during the forced reconnection. It displays the dynamics of a zoomed region as shown in the white-dotted box in Fig.2a. The first animation corresponds to the larger field-of-view as displayed in the Fig2a.  Panel $'$a$'$ is the composite image of AIA 171 \AA, 304 \AA~showing off-limb region.  Panel $'$b$'$ is a zoomed view of the {ROI} as shown by the dotted-white box in panel $'$a$'$. Panel $'$c$'$ is the schematic showing the formation of X-point in the far off-limb corona.}
\end{figure*}

In the present paper, using multi-wavelength observations of the solar corona from the Atmospheric Imaging Assembly (AIA) onboard the Solar Dynamics Observatory (SDO) on 3 May 2012, we establish directly that forced reconnection at a considerably high rate can occur locally in its magnetized plasma. It triggered in the corona when two oppositely directed magnetic field lines forming an X-point are perturbed by an external disturbance. This type of reconnection has only been reported in theory (Jain et al.,2005; Potter et al., 2019), and has never been directly observed in the Sun's large-scale corona. Although, as mentioned above an indirect signature of forced reconnection is claimed in the highly dynamic solar chromosphere at small spatial-scales for the release of tiny microflares (Jess et al.,2010). In Sect.~2, we present the observational data and its analyses. The observational results are described in Sect.~3. Sect.~4 outlines the numerical simulation, related setup, and results. The Discussion and Conclusions are depicted in the last section.

\begin{figure*}
\begin{center}
\hspace{-1cm}
\includegraphics[scale=0.6,angle=0,width=17cm,height=20.0cm,keepaspectratio]{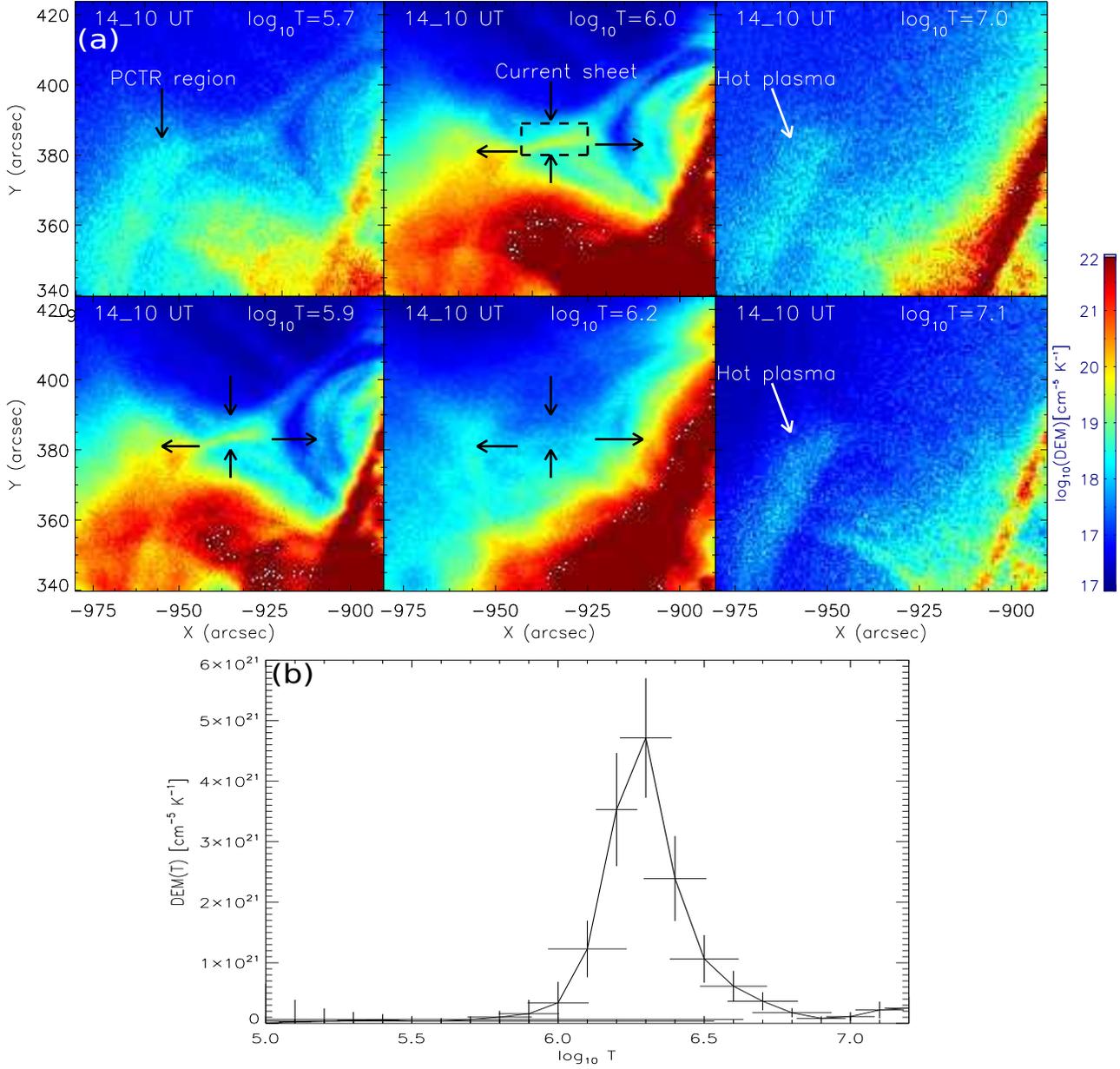}
\end{center}
\caption{The Differential Emission Measure (DEM) (Hannah \& Kontar, 2012) maps for different temperature bins between Log T$_{e}$ =5.0 and Log T$_{e}$ =7.1 (Fig.~3a) display that a prominence brings the overlying southward coronal magnetic field lines towards oppositely directed northward fields, and force externally the formation of an inflow at a X-point to trigger the forced reconnection. The estimated average temperature of the current sheet (Fig.~3a, black box region) using Hannah 
\& Kontar (2012) DEM technique peaks at log T$_{e}$=6.4 (Fig.~3b). The online animation (Fig3a-anim.mp4) shows the complete multi-temperature view of the observed forced reconnection. This animation runs during 13:30 to 14:45 UT.}
\end{figure*}
\section{Observational Data and Its Analyses}
 We analyze the data from the Atmospheric Imaging Assembly (AIA: Lemen et al. 2012) onboard the Solar Dynamic Observatory (SDO). The Solar Dynamic Observatory consists of three instruments, the Atmospheric Imaging Assembly (AIA), the Heliospheric Magnetic Imager and Extreme Ultraviolet Variability Experiment (EVE). The Atmospheric Imaging Assembly (AIA) has seven extreme ultraviolet (94, 131, 171, 193, 211, 304, 335 {\AA}), two ultraviolet (1600, 1700 {\AA}) and a visible (4500{\AA}) full disc imager with the 1.5 arcsec spatial resolution with a pixel size of 0.6 arcsec. It possess 12 s temporal resolution.
 We use the EUV temporal image data of AIA for our analysis. We have selected 2 hour time sequence data starting from 13:00 UT on 3$^{rd}$ May 2012 to observe a limb prominence. The basic calibration and normalization of the data were performed by using the Solarsoft IDL routine "aia\_prep.pro". \\
We use two channels of SDO/AIA data to analyze the prominence dynamics. AIA 304 {\AA} is dominated by the He II lines formed between (5-8)$\times 10^{4}$ K and 171 {\AA} is formed around (6-8)$\times 10^{5}$ K. The prominence flows are best observed in 304 {\AA} and the overlying hotter plasma and magnetic field regions in the 171 {\AA}. The composite images are constructed by combining the AIA 304 {\AA} and 171 {\AA} to observe the behavior of cooler prominence plasma and hotter field regions simultaneously.\\

In order to understand the thermal structures of the plasma flows, we obtain the Differential Emission Measure (DEM). We map the DEM with the different temperature from the six AIA filters, i.e, 94 {\AA} 131 {\AA}, 171 {\AA}, 193 {\AA}, 211 {\AA} and 335 {\AA}. We have used the method of Hannah \& Kontar (2012) to measure the differential emission from the prominence and surrounding magnetic field. The total emission measure from the given temperature interval is EM$_{T}$=DEM(T).$\Delta$T=$\int n_{e}^{2}dl$, which indicates that the amount of plasma integrated along the LOS over the per unit area. Here n$_{e}$ is an electron number density associated with temperature bin size $\Delta$T(K). It is an automated method, which returns a regularized DEM as a function of temperature (T). For the inversion, we use zeroth-order
regularization in the temperature range of log T(K)=5.0 to log T(K)=7.3 with 23 temperature bins at ∆log
T(K)=0.1 intervals. For the given six AIA filters, we obtain the DEM for the {ROI} in
the selected temperature regions (Fig.~3a). The DEM temperature map shows similar behavior as observed
into the AIA filters. From the DEM analysis, we observe that the cool and dense prominence plasma is surrounded by prominence-corona transition region (PCTR) and associated coronal magnetic field. The associated plasma is lying at a cool temperature (log T(K)=5.8). The oppositely oriented field regions, current sheet and externally driven plasma inflow as well as outflow with a certain time lag, are observed in the range of log T(K)=5.9-6.1, which corresponds to AIA 171 {\AA} filter.

\begin{figure*}
\begin{center}
\hspace{-0.8cm}
\includegraphics[scale=0.7,angle=90,width=17.cm,height=16.5cm,keepaspectratio]{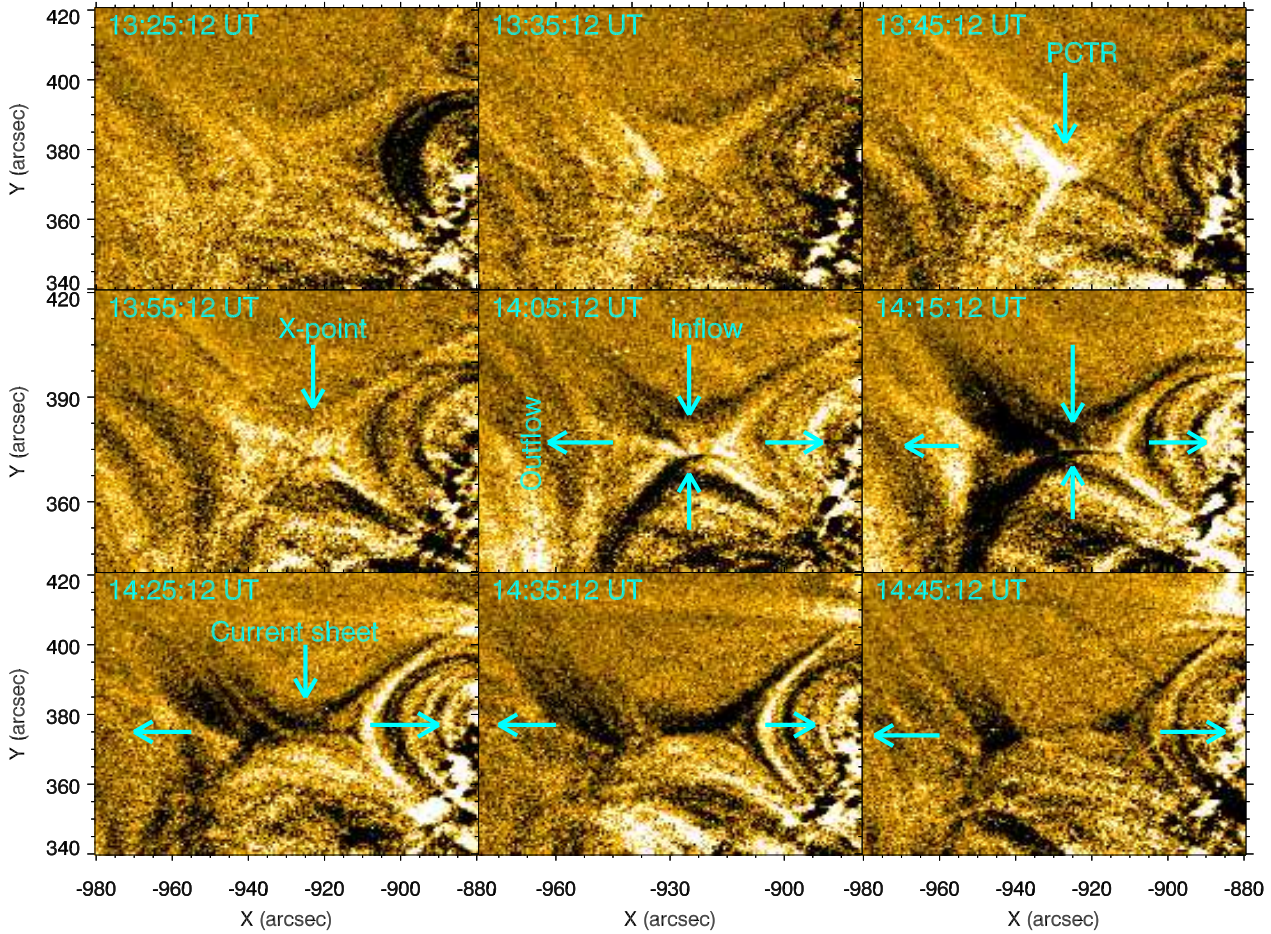}
\end{center}
\caption{\small \small Formation of X-point, and development of forced reconnection, and inflow/outflow regions are depicted in this figure. 
The running difference image sequence in the SDO/AIA 171 \AA~filter displays the motion of the prominence corona-transition region (PCTR) 
interface, which brings the southward branch of the coronal magnetic field towards the temporarily evolved X-point around 13:55 UT. 
The forced reconnection is subsequently triggered thereafter the initial onset of externally driven inflows. Plasma inflow occurs from the north-south direction (cf.Fig.,~5$'$b$'$, left middle image panel at 13:48 UT), 
and after the reconnection, the outflow is seen in the east-west direction (cf., Fig.,~5$'$c$'$; left lower panel during 13:57-14:45 UT). A dynamic current sheet is also formed during the forced reconnection process. The animation $'$fig4-anim.mp4$'$ shows the annotated motion of the prominence, plasma inflows at X-point, forced reconnection, and plasma outflows as seen in the difference image sequence of AIA 171 \AA~. This movie runs during the time between 12:50 UT to 14:59 UT.}
\end{figure*}

\section{Observational Results}
A prominence eruption has been observed using SDO/AIA 171 $\AA$ EUV running difference image data on 3$^{rd}$ May 2012 (Fig.~1). A spiral-shaped structure (visible in AIA 171 \AA, but not in AIA 304 \AA) has erupted with spherical top and conical base covered by coronal loops. Initially, a C1.6 class solar flare is initiated at 13:24 UT and it may act as trigger for the prominence eruption (Fig.~1). The eruptive prominence has conical base and it is expanded spatially and temporally. The expansion of the eruptive prominence perturbs the surrounding coronal field mostly in far northward regions, where the X-point configuration is temporarily created in the space. The eruptive prominence associated coronal field (PCTR) is expanded and forced the southward segment of the coronal field lines to move towards creating the X-point and enabling the forced reconnection. The southward segment is associated with a prominence. In the present scenario, the two branches of magnetic field lines are visible in the solar corona (cf., the white-dotted box in Fig.~2$'$a$'$). The southward branch of the magnetic field envelopes a cool prominence, and is part of a large-scale closed field lines structure in the corona. The northward branch of the magnetic field is anchored at the limb, while its other end is open in the diffused corona. 
These two branches of large-scale coronal fields are separated by a low lying closed loop system with both foot-points anchored at the limb.
The schematic (Fig.~2$'$c$'$) 
clearly outlines the observed magnetic field configuration where the southward field, plus an embedded prominence (green) and open fields (blue) lie at the either sides of a low lying coronal loop system (red). The southward branch of the magnetic field lines is driven by the embedded prominence, moving towards the northward branch to form a temporary X-point.
An inflow region is initially established at the X-point to enable the forced reconnection (see animation $'$fig2a-anim.mp4$'$ and $'$fig2b-anim.mp4$'$). We do not identify the X-point based on the magnetic field data as the observed region is far from the limb. Instead, we introduce it in the multi-wavelength imaging observations while we observe the junction point of two opposite sets of the magnetic field lines separated by a loop system (Figs.2$'$a$'$, $'$b$'$). 

The DEM map at different temperature bins between
Log T$_{e}$=5.0 and Log T$_{e}$=7.3 (Hannah \& Kontar 2012; Fig.~3$'$a$'$) and associated time-sequence (see animation $'$fig3a-anim.mp4$'$; duration: 13:30 UT to 14:45 UT on 3 May 2012) also show that the cool prominence is enveloped by the southward branch of the hot and magnetized coronal plasma. 
The prominence plus the associated coronal field moves northward forming an X-point along with the northward branch of the magnetic fields. This enables first an inflow towards the evolved X-point in the corona and consequently forced magnetic reconnection begins. A dynamic current sheet is also formed during the reconnection process. 
We have introduced the plot related to the evolution of temperature, which is estimated at the reconnection site using the method of Hannah \& Kontar (2012). The estimated average temperature peaks at log T(K)= 6.4 in the current sheet region (Fig.~3$'$b$'$). Xue et al. (2018) have also reported the formation of comparatively cooler current sheet observed in the transition region maintained at typical coronal temperature (1.6 MK) when it is driven by the small scale sunspot in the photosphere. The evolution of temperature at the reconnection site and current sheet should be the characteristics of the normal magnetic reconnection in form of local heating and flare activities. However, in the present observations, we observe the prominence driven forced magnetic reconnection. There is no flare activity  associated with this forced magnetic reconnection, so we may not observe the evolution of high temperature and related energy release there at the current sheet (e.g., Chen et al. 2001; Chen \& Ding 2006; Jess et al. 2010). Therefore, the prominence driven forced magnetic reconnection is at work in the present observational base-line, however, we do not observe the evolution of extremely hot plasma there. Chen \& Ding (2006) have concluded that the high densities found in the photosphere and chromosphere, coupled with a partially ionized and highly collisional plasma. Therefore, the energy release through the magnetic reconnection may be consumed by these highly dense and collisional plasma. In the present work, we observe that the  prominence driven coronal field lines reach to the reconnection site at 13:55 UT. Mostly PCTR as visible in 171 {\AA} AIA channel is involved in the present reconnection scenario which is enveloping cool and denser prominence plasma. The prominence plasma is characterized as a partially ionized, dense and collision dominated plasma. Therefore, we conjecture that the energy produced in the forced magnetic reconnection may be consumed in  form of  the kinetic energy of the outflowing plasma rather heating it.\\

The sequence of difference images (Fig.~4; see $'$fig4-anim.mp4$'$) demonstrates that in the localized corona the inflow and associated X-point is created externally by the motion of the prominence around 13:55 UT. The hot plasma in the prominence corona-transition region (PCTR) also drifts towards the X-point, and brings the southward branch of the magnetic field near the inflowing northward magnetic field, triggering the forced reconnection. The bi-directional inflow starts from the north-south direction towards the X-point around 13:48 UT (Figs.~4, 5$'$b$'$), and forced reconnection is triggered. Plasma outflows are also seen in the perpendicular east-west direction of the inflows along with the formation of a dynamic and elongated current sheet. It should be noted that the externally driven plasma inflows start first at $\approx$ 13:48 UT (Figs.~4, 5$'$b'$'$), and after the reconnection the outflows start. This is the key observational aspects of the forced reconnection. The time-distance maps (cf., Figs.~5$'$b$'$, $'$c$'$) along the slits $'$S$_{1}$$'$ (North-South direction), \& $'$S$_{2}$$'$  (East-West direction) near the X-point (cf., Figs.~5$'$d$'$) show that the bi-directional inflow is followed by plasma outflow at the X-point. 

We have shown the inflow and outflows velocities in Figs.~5$'$b$'$, $'$c$'$ and discussed the whole complex dynamical processes that occurred near the X-point. We observe that bi-directional inflow and outflow occurred near the reconnection point. In the previously reported results (e.g., Savage et al. 2012; Takasao et al. 2012; Su et al. 2013; Xue et al. 2016; Yan et al. 2018 references their cite in), the bi-directional inflows and outflows were moving with different velocities during the onset of the magnetic reconnection. We clearly emphasize the physical significance of the detected inflows and outflows in the present observational base-line in a more detailed manner under the frame-work of a case study of the forced reconnection (Figs. $'$2-4$'$). In the present case, plasma motions are more complex. Along  slit $'$S$_{2}$$'$, tangled plasma motions are driven most initially by prominence and they try to bring the different sets of magnetic field lines towards the X-point location (Fig.~5$'$c$'$; see the green-dashed lines). Their almost constant projected speed (8-9 km s$^{-1}$) indicates that different set of field lines are driven uniformly by an external driver, which is a prominence in the present case. Based on their spatial locations, some coronal field lines penetrate deeper towards the X-point, while some other remain to the outer-periphery (see various green-dashed paths in panel $'$c$'$ of Figure 5).  It is visible in  panel $'$c$'$ of Figure 5 that as the prominence driven forced motions start, an another set of inflow motions of the magneto-plasma system is also initiated typically from North-South direction along slit $'$S$_{1}$$'$ with the speed of 3-7 km s$^{-1}$.  Prominence driven motions are tangled w.r.t East-West direction, while the another set of inflows were having a preferred North-South direction of the motions. However, in principle, they collectively create a temporary X-point geometry in the localized corona where reconnection is eventually forced. While, the North-South directed inflows were continuous even after the commencement of the forced reconnection around $\approx$13:54 UT, the  reconenction generated outflows preferentially started in the East-West direction with the re-orientation and re-joining of the various sets of magnetic field lines (Figure 4 and Movie 2). The field lines that reached most closer to the X-point (shown by a red arrow in panel $'$c$'$ of Figure 5), they exhibit an outflow (11 km s$^{-1}$) in the west direction along slit $'$S$_{2}$$'$, From this point an another set of outflowing plasma (5 km s$^{-1}$) is originated in the East direction along slit $'$S$_{2}$$'$. Remaining set of the magnetic field lines and associated frozen-in plasma, which could not reach near the typical X-point (see various green-dashed lines in panel $'$c$'$ of Figure 5), they started out-flowing in West during the reconnection from their most nearest approach to the reconnection site (temporary X-point). This whole description is itself able to conjecture that the magnetic reconnection process in the present observational base-line is a forced reconnection. 

In conclusion, we have explained above the dynamics along the slit $'$S$_{1}$$'$ and $'$S$_{2}$$'$ respectively in Figs. 5b and c. It is visible in panel $'$c$'$ of Fig. 5 that as the prominence driven magnetic structures move, and it drives various segments of prominence-corona-Transition Region (PCTR) towards the X-point. This plasma dynamics starts earlier around 13:20 UT due to an external forcing by the prominence. Once this dynamics was developing from the southward direction, the plasma inflow motions also start along slit $'$S$_{1}$$'$ from North-South direction. These motion of the magnetic field lines start at $\approx$13:48 UT (vertical voilet-dotted line in Fig. 5b), which acts as an another set of inflow the inflow along the slit $'$S$_{1}$$'$. These inflows along ($'$S$_{1}$$'$ \& $'$S$_{1}$$'$) brings two sets of magneto-plasma system towards temporarily formed X-point. Reconnection at the X-point is started around $\approx$13:54 UT (vertical blue-line in Fig. 5c). The plasma motions continue from North-South direction even after the commencement of the reconnection at the X-point, which shows the reduction of the width of reconnection region and thinning of the associated current sheet from this direction.

Therefore, we consider the pair of inflow and outflow due to forced reconnection which start close to the X-point. There is a time lag of a few minutes between inflows and outflows that occur in the observed forced reconnection region at the X-point. The intensity at the X-point firstly peaks in the cool AIA 304 \AA~filter (Log T$_{e}$=4.7) at $\approx$13:50 UT, which indicates that PCTR plasma segment first comes inside to form the X-point for forcing the reconnection. Once reconnection begins, the emission in the high temperature AIA filters 171 \AA~(Log T$_{e}$=5.9), and 131 \AA~(Log T$_{e}$=5.6, 6.9) peak between $\approx$14:00 UT-14:05 UT. This exhibits a quick energy release at the X-point in $\approx$10 min time scale. In the same duration the intensity from the higher temperature filter AIA filter 193 \AA~(log T$_{e}$=6.2, 7.1) and 211 \AA~(log T$_{e}$=6.4) is minimum. Intensities of these filters rises between 14:05-14:12 UT, which indicates the spontaneous energy release near the X-point (cf., Fig.~5a) during the reconnection process after its onset. Later, secondary post-reconnection processes are established at the X-point (e.g., dynamics and stretching of current sheet, outflows, cooling). The most significant aspect of this observed forced reconnection is its high rate (V$_{inflow}$/V$_{outflow}$) ranging between $\approx$0.15-0.27. A reconnection rate, as a dimensionless quantity, can be approximated by the ratio of inflow to the outflow speed (Priest \& Forbes, 2007). We have estimated the reconnection rate using the inflow and outflow velocity along the slit S1 and S2. The magnetic reconnection rate estimated as the ratio of inflow and outflow velocity along the current sheet, i.e, M= V$_{inflow}$/V$_{outflow}$ (cf., Priest 2014; Xue et al. 2015).
\begin{figure*}
\begin{center}
\hspace{-1cm}
\includegraphics[scale=0.6,angle=0,width=17cm,height=17.0cm,keepaspectratio]{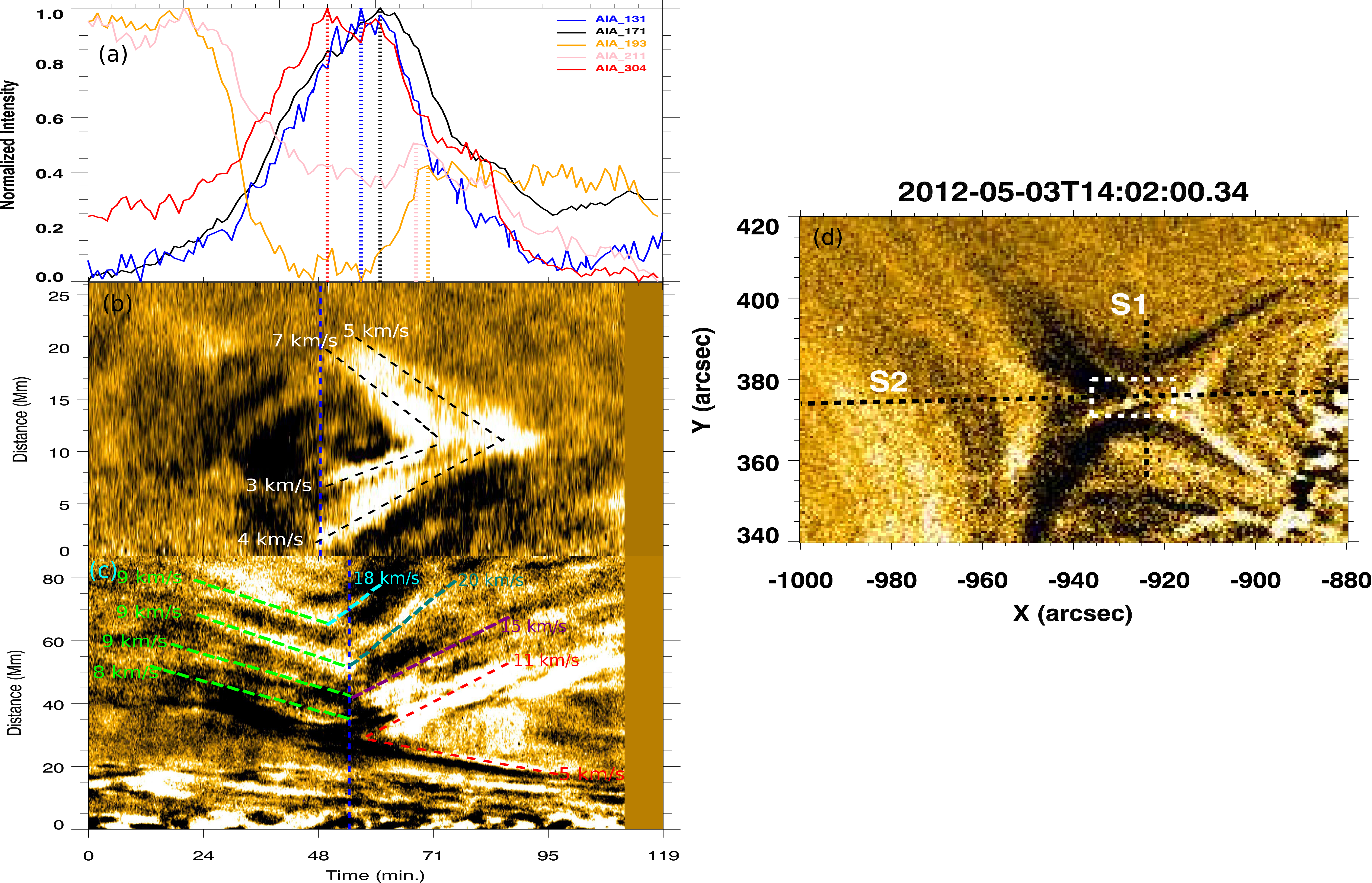}
\end{center}
\caption{\small \small The analyses of inflows, outflows, and nature of EUV emissions at the reconnection site are described in this figure.
The temporal variation of EUV plasma emissions (304 \AA, 211 \AA, 193 \AA, 171 \AA, 131 \AA~) has been measured to understand qualitatively the thermal response of the reconnection site (left top panel $'$a$'$). The slit $'$S$_{1}$$'$ and $'$S$_{2}$$'$ are analyzed for inflow and outflow plasma motions ($'$d$'$) across the X-point. These slits have been selected to deduce the distance-time maps ($'$b, c$'$) for estimating their speeds close to the X-point. }
\end{figure*}
\section{Numerical Modeling of the Forced Reconnection}

We develop a physical model of the observed forced reconnection by taking a weakly diverging, current-free equilibrium magnetic field embedded in a hydrostatic magnetohydrodynamic (MHD) solar atmosphere with an appropriate temperature profile (Vernazza et al., 1981; Jel\'inek et al., 2015). We used an MHD code, known as FLASH (Fryxell et al., 2000) to simulate the observed forced magnetic reconnection. 

\subsection{Governing MHD equations}
The numerical model for the forced reconnection usually makes use of a gravitational-stratified solar atmosphere. The plasma dynamics is investigated using a set of 2-D, time-dependent, and  resistive MHD equations for the fully ionized plasma. We use the FLASH code (Lee, 2013; Fryxell et al., 2000) for the simulation of the plasma  dynamics at an X-point in the model solar corona, where two oppositely directed magnetic field lines meet. The governing MHD equations are written in their conservative forms, which are outlined as follows (Priest, 2014):
\begin{equation}\label{eq1}
\frac{\partial \varrho}{\partial t} + \nabla \cdot(\varrho \bm{\mathrm{v}}) = 0,
\end{equation}
\begin{equation}\label{eq2}
\frac{\partial \varrho \bm{\mathrm{v}}}{\partial t} + \nabla \cdot \left(\varrho \bm{\mathrm{v}} \bm{\mathrm{v}} -\bm{\mathrm{B}}\bm{\mathrm{B}}\right) + \nabla p_{*}=\varrho \bm{\mathrm{g}},
\end{equation}
\begin{eqnarray}\label{eq3}
\frac{\partial \varrho E}{\partial t} + \nabla \cdot \left[\left(\varrho E+p_{*}\right)\bm{\mathrm{v}} - \bm{\mathrm{B}}(\bm{\mathrm{v}} \cdot \bm{\mathrm{B}})\right] =  \nonumber \\
= \varrho \bm{\mathrm{g}} \cdot \bm{\mathrm{v}} + \nabla \cdot (\bm{\mathrm{B}} \times (\eta \nabla \times \bm{\mathrm{B}})),
\end{eqnarray}
\begin{equation}\label{eq4}
\frac{\partial \bm{\mathrm{B}}}{\partial t} + \nabla \cdot (\bm{\mathrm{v}}\bm{\mathrm{B}}-\bm{\mathrm{B}}\bm{\mathrm{v}})=-\nabla \times (\eta \nabla \times \bm{\mathrm{B}}),
\end{equation}
\begin{equation}\label{eq5}
\nabla\cdot\bm{\mathrm{B}}=0.
\end{equation}
Here $\varrho$, $\bm{\mathrm{v}}$, $\bm{\mathrm{B}}$, $\bm{\mathrm{g}} =[0,-g_{\sun},0]$, $\eta$ are respectively mass density, flow velocity, magnetic field strength, gravitational acceleration, and magnetic diffusivity. 
We use $g_{\sun}$=$274~\mathrm{m s^{-2}}$ in the model atmosphere. The magnetic diffusivity (Priest, 2014) is taken constant throughout the numerical box and corresponds to the relation $\eta = 10^9 T^{-3/2}$ for the coronal temperature $T= 10^6~\mathrm{K}$. To establish the magnetic reconnection in our model corona with defined X-point at an appropriate rate, we consider magnetic diffusivity much more larger compared to the collisional diffusivity (Jel\'inek et al., 2017).

The pressure $p_{*}$ in the model solar atmosphere can be approximated as follows:
\begin{equation}\label{eq6}
p_{*} = \left(p + \frac{B^2}{2 \mu_0}\right),
\end{equation}
Here, $p$ is the thermal pressure of the plasma gas, $B$ is the coronal magnetic field.
Here, $E$ in Eq. (\ref{eq3}) is attributed as specific total energy, which is expressed by the following equation:
\begin{equation}\label{eq7}
E = \epsilon + \frac{v^2}{2} + \frac{B^2}{2 \mu_0 \varrho},
\end{equation}
where $\epsilon$ is the specific internal energy of the system which can be approximated as follows:
\begin{equation}\label{eq8}
\epsilon = \frac{p}{(\gamma-1)\varrho}.
\end{equation}
It consists of $\gamma$, which is known as adiabatic index and it attains a value $5/3$. The plasma flow velocity and its magnitude is represented by $v$. The magnetic permeability $\mu_0$ is having a value of $1.26 \times 10^{-6}~\mathrm{H m}^{-1}$ in the free space as the coronal plasma is approximated as a highly rarefied magnetized plasma gas.

%
%
\subsection{Numerical code and simulation setup}
%
We have solved 2-D time-dependent, resistive, non-linear MHD equations (1)-(4) numerically by implementing the stringent FLASH code. This utilizes 2$^{nd}$ and 3$^{rd}$ order Godunov solvers. These solvers consist of different slope limiters as well as Riemann solvers. It also uses the scheme of adaptive mesh refinement (AMR) (Chung, 2002). Multi-dimensional integration in the Godunov solver is performed by corner transport upwind method. On the other hand, preservence of the divergence free constraints on the coronal magnetic field in the Godunov solver is complied using the constrained transport algorithm, which further makes magnetic field completely source free. The minmod slope limiter and the Riemann solver are used in the Roe approximation (Toro, 2006) in the present simulation as embedded in the numerical code. The major benefits of AMR
technique is that this method refines a numerical grid with an exceptionally steep spatial profiles of various physical variables in the model atmosphere. It keeps the grid coarse over those places where the fine spatial resolution within the model solar atmosphere is not required in the measurements. In the present case, the AMR method acts on controlling the numerical errors in the proximity of mass density gradient in the flowing plasma cells. This exhibits a lowering of the numerical magnetic diffusion at different numerical cells. This is necessary to do the modelling of the forced magnetic reconnection.

For modelling the forced manetic reconnection, we choose a 2-D simulation box of height and width respectively of $(40, 40)~\mathrm{Mm}$. The spatial resolution of each numerical grid is estimated using the AMR method in the present numerical model. We make a similar setup as in the case of the vertical current-sheet in the magnetized model solar corona (Jel\'inek et al., 2015). The AMR grid with the min/max level of the refinement blocks is set to $3/7$. This constitutes $1285$ numerical blocks to make a model. As each block consits of $8 \times 8$ numerical cells, the total number of blocks corresponds to $82~240$ cells. The  achieved maximum spatial resolution is $3.9~\mathrm{km}$ in both directions in the 2-D simulation box mimicking the resistive model solar corona. We incorporate free boundary conditions, so the incident signals on the boundaries of the numerical box  may leave the it without any reflection from the boundaries.  This attributes in minimizing the numerical errors.

Prior to running the numerical simulations of the forced magnetic reconnection, we have examined that the model solar atmosphere remains in the static equilibrium for the adopted grid resolution. This is achieved by executing an initial simulation without implementing any velocity pulse, which signifies the existence of magnetostatic solar corona.

%
%
\subsection{Initial equilibrium solar atmosphere}
%
For a stationary ($\bm{\mathrm{v}} = \bm{0}$) equilibrium, the magnetic Lorentz force and gravitational forces are jointly balanced
by the plasma gas pressure gradient working opposite to them,
\begin{equation}\label{eq8}
-\nabla p+\bm{\mathrm{j}}\times\bm{\mathrm{B}} + \varrho \bm{\mathrm{g}} = \bm{\mathrm{0}}.
\end{equation}
Keeping the view of a force-free magnetic field ($\bm{\mathrm{j}} \times \bm{\mathrm{B}} = \bm{\mathrm{0}}$) at the X-point in the model corona, the
hydrostatic equations mimicking the initially stable atmosphere further yielded in the following form:
\begin{equation}\label{eq9}
p_{\rm h}(y) = p_0 \exp\left[-\int\limits_{y_0}^{y} \frac{1}{\Lambda(\tilde{y})}\mathrm{d}\tilde{y}\right],
\end{equation}
\begin{equation}\label{eq10}
\varrho(y) = \frac{p(y)}{g_{\sun}\Lambda(y)}.
\end{equation}
Here
\begin{equation}\label{eq11}
\Lambda(y) = \frac{k_\mathrm{B}T(y)}{\overline{m}g_{\sun}},
\end{equation}
is the scale-height of the pressure in the static corona that in the frame-work of an isothermal atmosphere exhibiting the vertical height over which it decreases by $1/e$ factor. Here $k_\mathrm{B} = 1.38 \times 10^{-23}~\mathrm{J\cdot K^{-1}}$ is the Boltzmann constant. The mean particle mass is given as $\overline{m} = 0.6\, m_\mathrm{p}$ where $m_\mathrm{p} = 1.672 \times 10^{-27}~\mathrm{kg}$ is the mass of the proton mass.

The solenoidal (divergence free) magnetic field condition, $\nabla\cdot\bm{\mathrm{B}}=0$, is universally valid condition that uses a magnetic flux function, $\bm{A}$, which is given as follows:
\begin{equation}\label{eq12}
\bm{\mathrm{B}} = \nabla \times \bm{\mathrm{A}}.
\end{equation}
We construct the non-potential X-point (a null-point) as follows (Jel\'inek et al., 2015; Parnell et al., 1997):
$$\bm{\mathrm{A}} = [0,0,A_z]$$ 
with 

\begin{equation}\label{eq13}
A_z = \frac{1}{4} B_{\mathrm{0}} [(\mathcal{J}_t - \mathcal{J}_z) y^2 - (\mathcal{J}_t + \mathcal{J}_z) x^2].
\end{equation}
Here $\mathcal{J}_t$ is the threshold current depending on the parameters associated with the potential magnetic field in the solar corona.
It is taken as a constant in our numerical calculations. The parameter $\mathcal{J}_z$ is the magnitude of the current which is perpendicular to 
the plane of the X-point (a null-point) as constituted in the model corona. Both $\mathcal{J}_t$ and $\mathcal{J}_z$ are free parameters, which establish the configuration of the magnetic field and its shape. In the present work it is in form of the X-shaped curent sheet along with a null point (Pernell et al., 1997).

The gas pressure and mass density are estimated as follows in the initial stable solar atmosphere (Jel\'inek et al., 2017; Solov'ev, 2010):
\begin{equation}\label{eq14}
p(x,y) = p_{\rm h} - \frac{1}{\mu_0}\left[\int\limits_{-\infty}^{x}\frac{\partial^2 A}{\partial y^2}\frac{\partial A}{\partial x}\mathrm{d}x +
\frac{1}{2}\left(\frac{\partial A}{\partial x}\right)^2\right],
\end{equation}

\begin{eqnarray}\label{eq15}
\varrho(x,y) = \varrho_{\rm h}(y) &+& \frac{1}{\mu_0 g_{\sun}}\Bigg\{\frac{\partial}{\partial{y}}\Bigg[\int\limits_{-\infty}^{x} \frac{\partial^2 A}{\partial y^2}\frac{\partial A}{\partial x}\mathrm{d}x + \nonumber \\
&+& \frac{1}{2}\Bigg(\frac{\partial A}{\partial x}\Bigg)^2\Bigg] - \frac{\partial A}{\partial y} \nabla^2 A\Bigg\}.
\end{eqnarray}
By using Eq.~(\ref{eq13}) in these general formulae of various physical parameters, we find the expressions for the gas pressure (Jel\'inek et al., 2015)
\begin{equation}\label{eq16}
p(x,y) = p_h(y) - \frac{B_0^2}{4\mu_0} \mathcal{J}_z (\mathcal{J}_t + \mathcal{J}_z) x^2,
\end{equation}
and the mass density in the equilibrium solar atmosphere
\begin{equation}\label{eq17}
\varrho(x,y) = \varrho_h(y) + \frac{B_0^2}{2\mu_0 g} \mathcal{J}_z(\mathcal{J}_t - \mathcal{J}_z) y.
\end{equation}

The numerical methodology and description of the FLASH code used in the present work is adopted and outlined from various detailed descriptions (Jel\'inek et al., 2015, 2017).

\section{Numerical Results}

The X-point is created in the equilibrium atmosphere along with an appropriate inclusion of the resistivity (Fig.~6). The hydrostatic plasma equilibrium is perturbed by a localized velocity pulse set at the left side of the X-point, which resembles an external disturbance forcing the reconnection. The disturbance mimics an effect of a velocity field created by the moving prominence as seen in the observations. It is given as follows: 
\begin{equation}\label{eq22}
v_x = -A_0\cdot \frac{x}{\lambda_y} \cdot
\exp{\left[-\frac{(x-L_{\mathrm{P}})}{\lambda_x}\right]^2}\cdot
\exp{\left[-\frac{y}{\lambda_y}\right]^2},
\end{equation}
Here $\lambda{_x}$= 0.1 Mm and $\lambda{_y}$= 0.1 Mm are the width of the external disturbance ($v_x$), which is a symmetric Gaussian-shaped velocity pulse in the vicinity of a null-point (Fig.~6).
The $x$= 0.2 Mm and $y$= 15 Mm are the positions of the disturbance in the model corona. The amplitude $A_{o}$ is fixed as 0.1 Mm s$^{-1}$.

\begin{figure*}
\begin{center}
\mbox{
\hspace{-2.5cm}
\includegraphics[scale=0.6,angle=0,width=4.3cm,height=4.8cm,keepaspectratio]{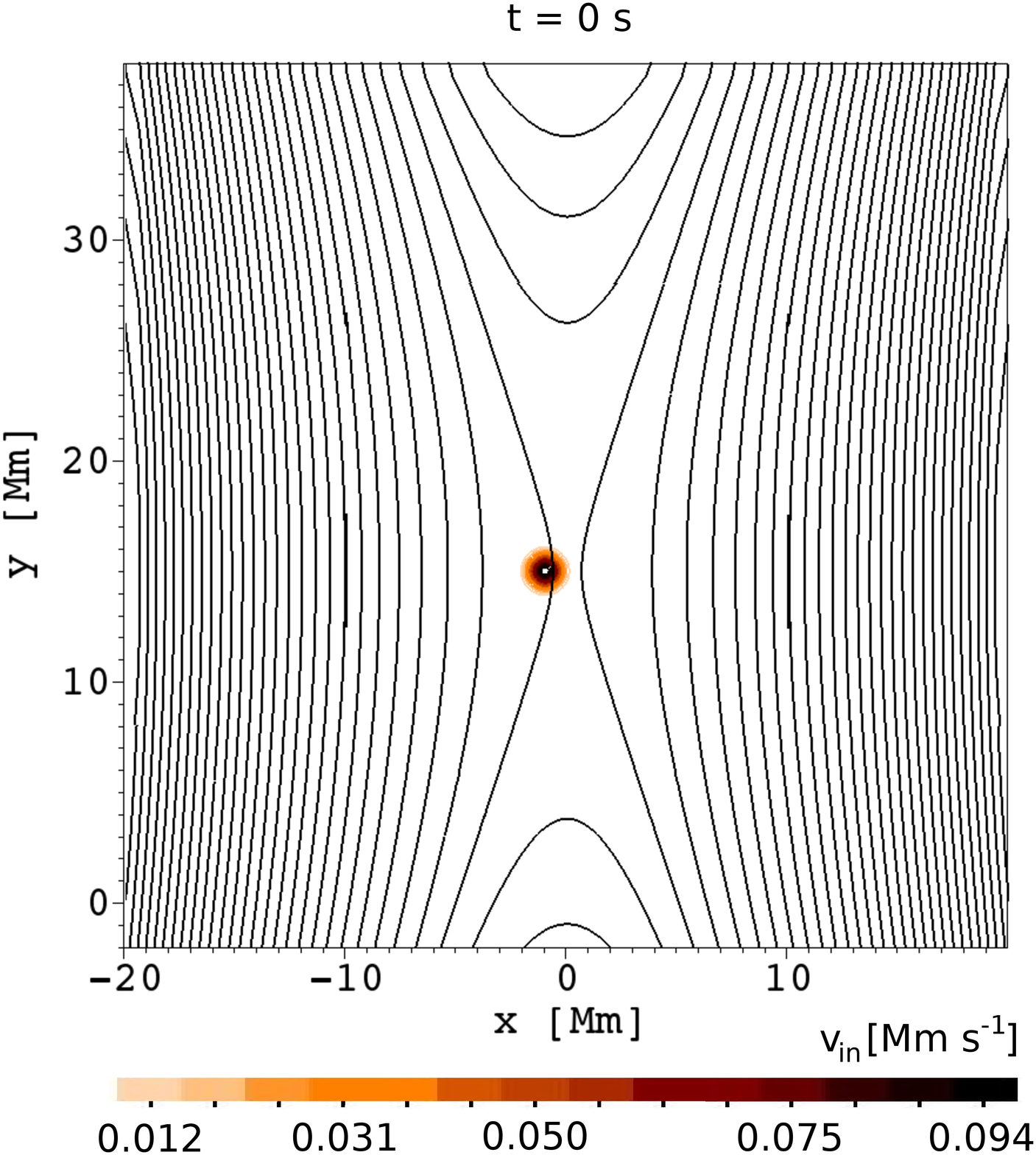}
\hspace{-0.42cm}
\includegraphics[scale=0.6,angle=0,width=4.3cm,height=4.8cm,keepaspectratio]{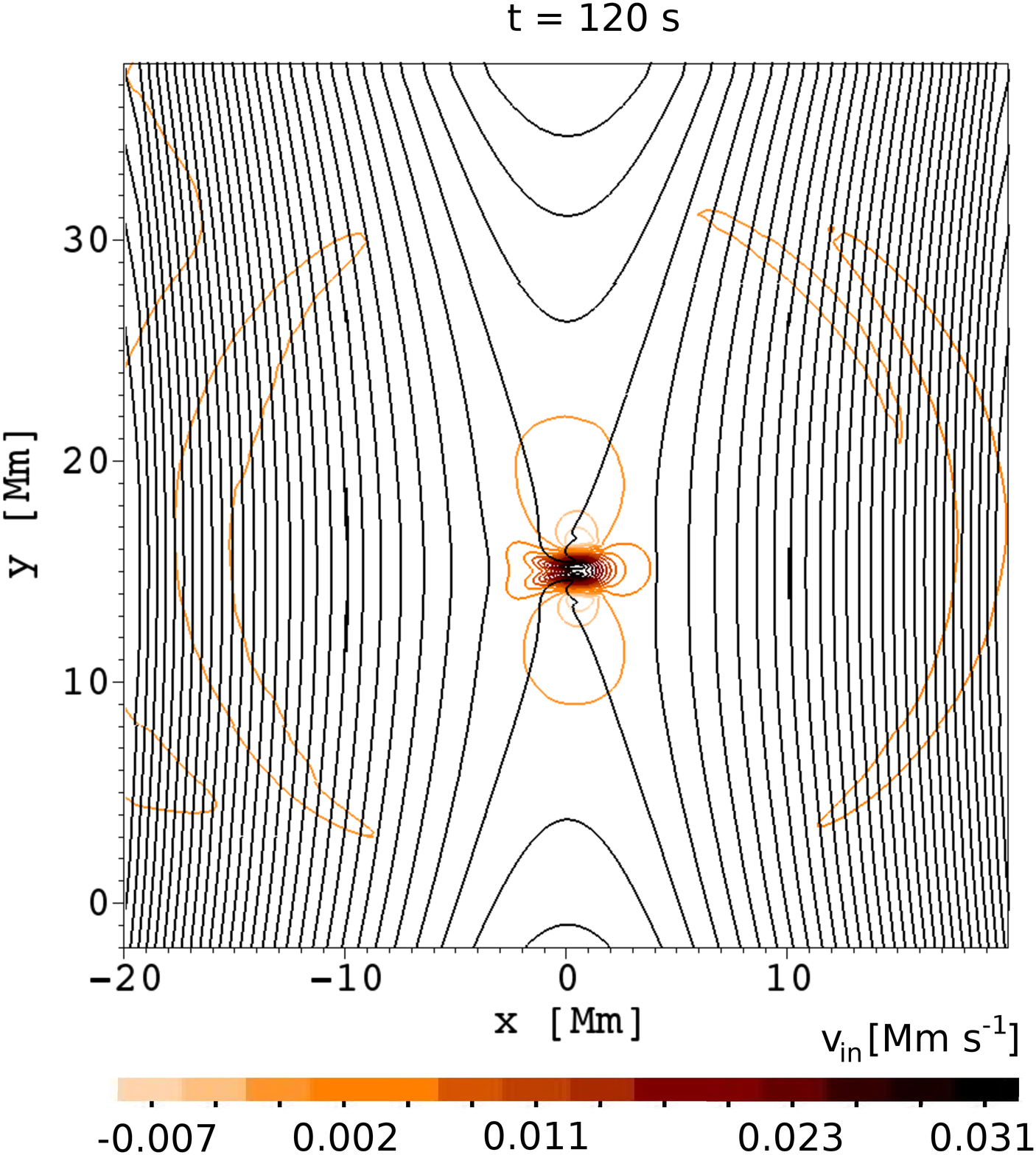}
\hspace{-0.42cm}
\includegraphics[scale=0.6,angle=0,width=4.3cm,height=4.8cm,keepaspectratio]{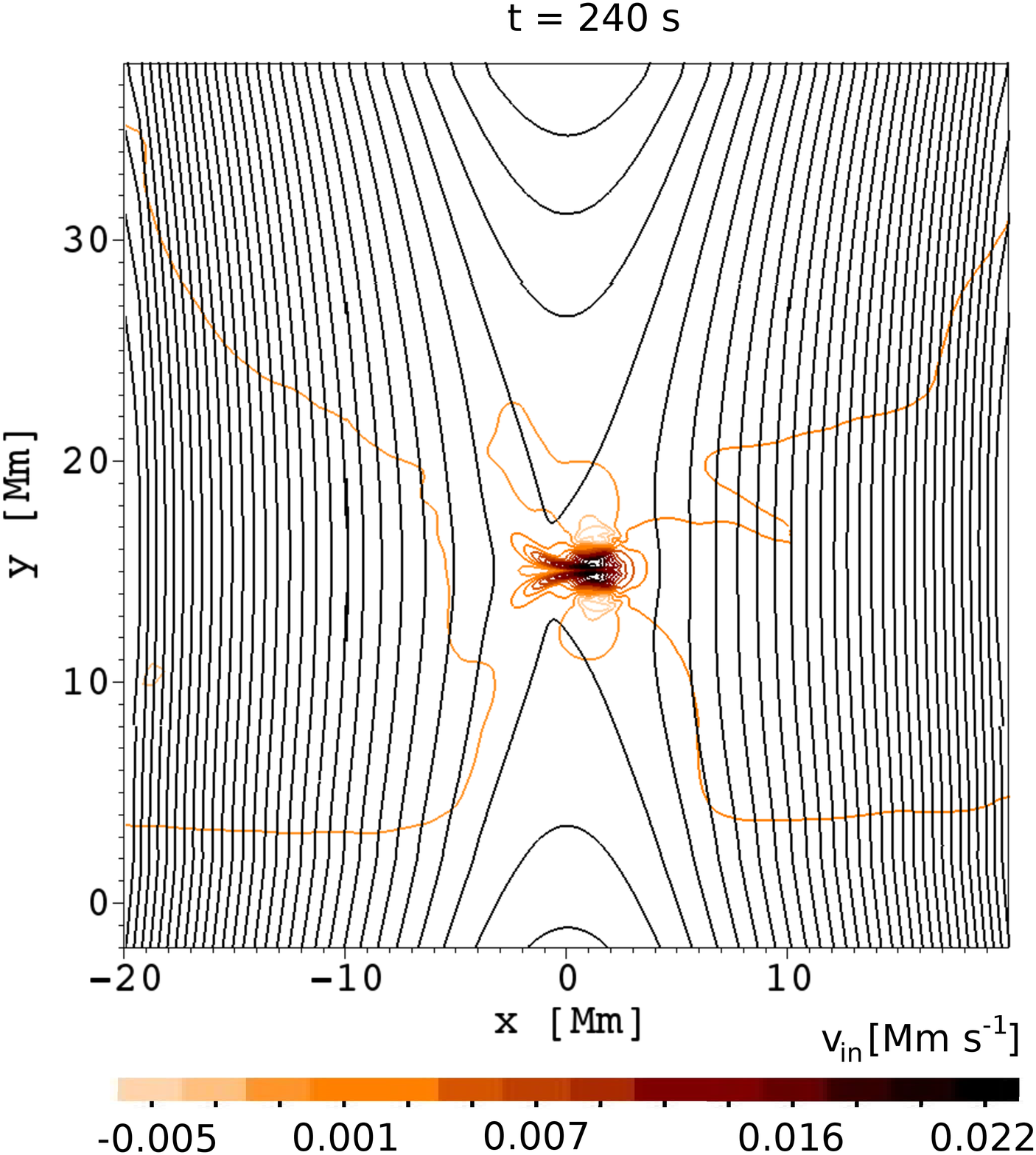}
\hspace{-0.42cm}
\includegraphics[scale=0.6,angle=0,width=4.3cm,height=4.8cm,keepaspectratio]{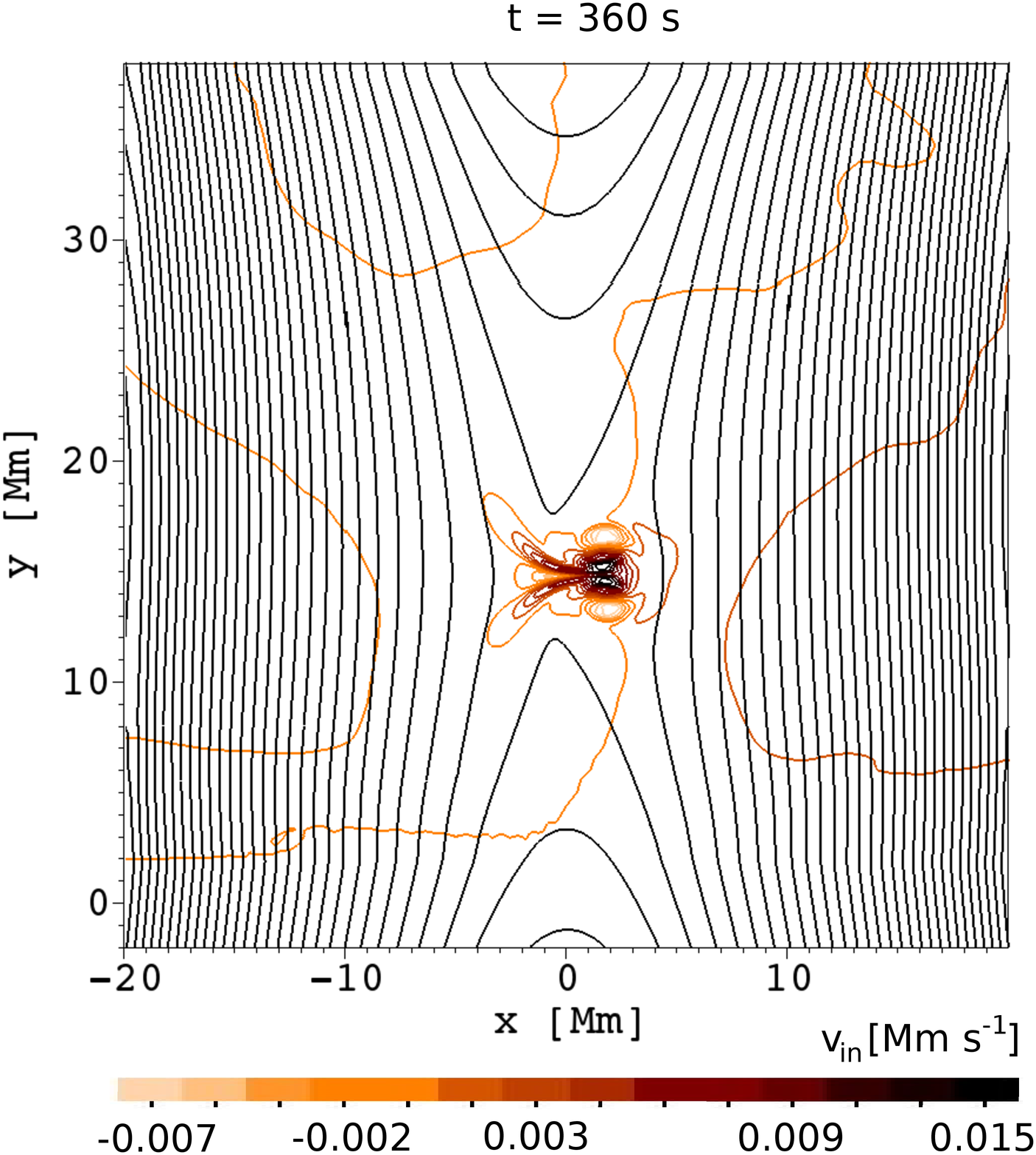}
\hspace{-0.42cm}
\includegraphics[scale=0.6,angle=0,width=4.3cm,height=4.8cm,keepaspectratio]{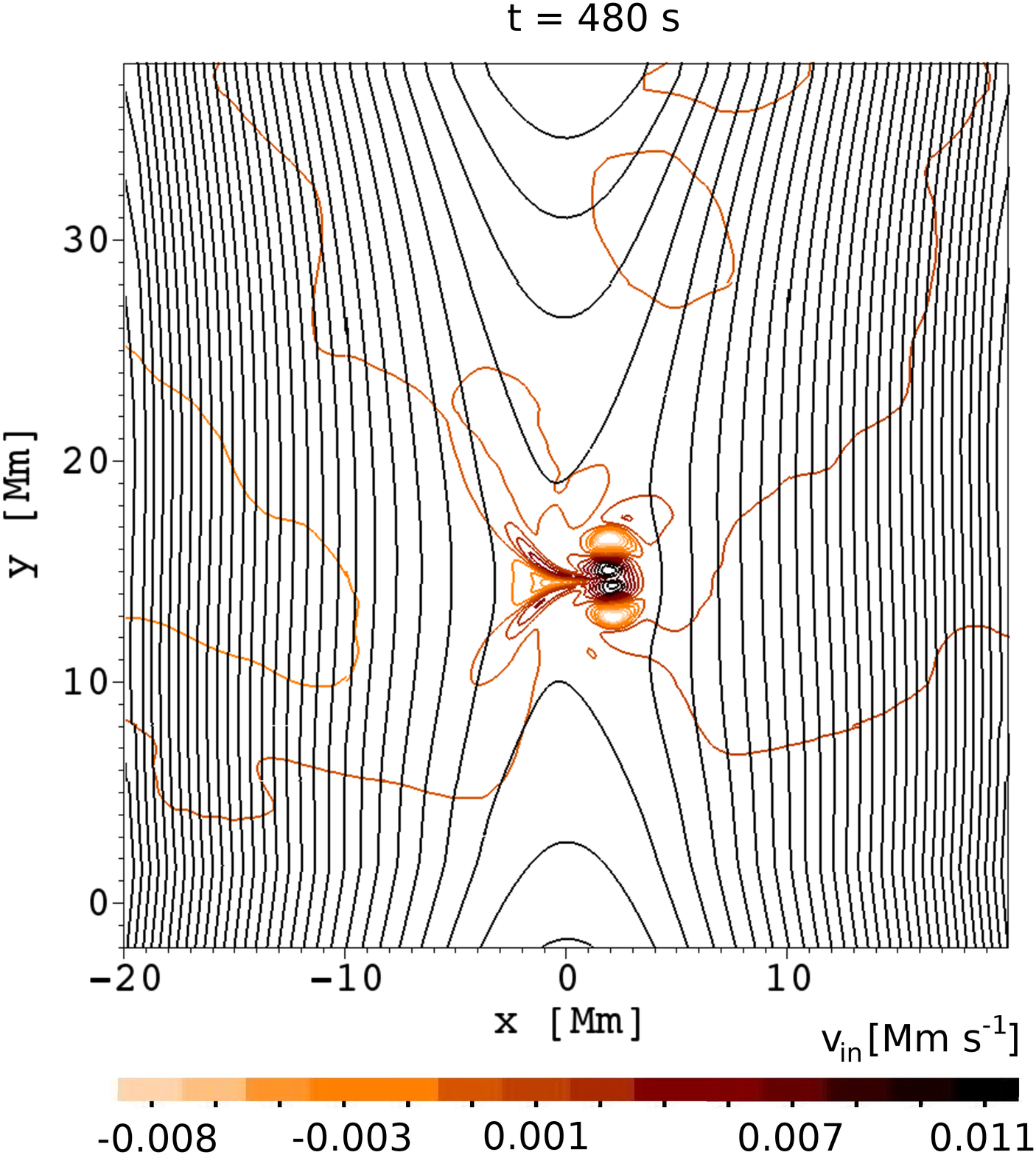}
}
\mbox{
\hspace{-2.5cm}
\includegraphics[scale=0.6,angle=0,width=4.3cm,height=4.8cm,keepaspectratio]{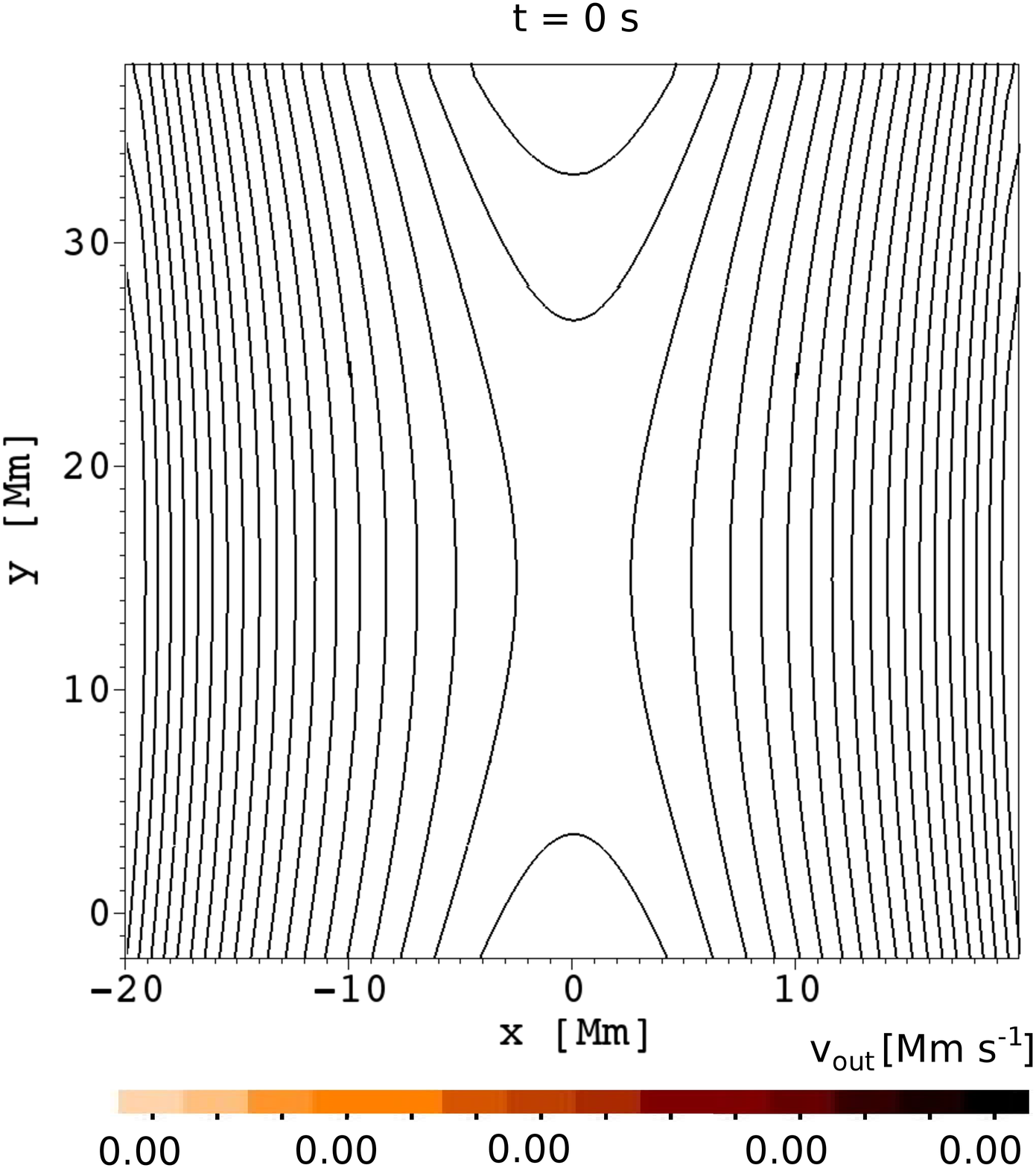}
\hspace{-0.42cm}
\includegraphics[scale=0.6,angle=0,width=4.3cm,height=4.8cm,keepaspectratio]{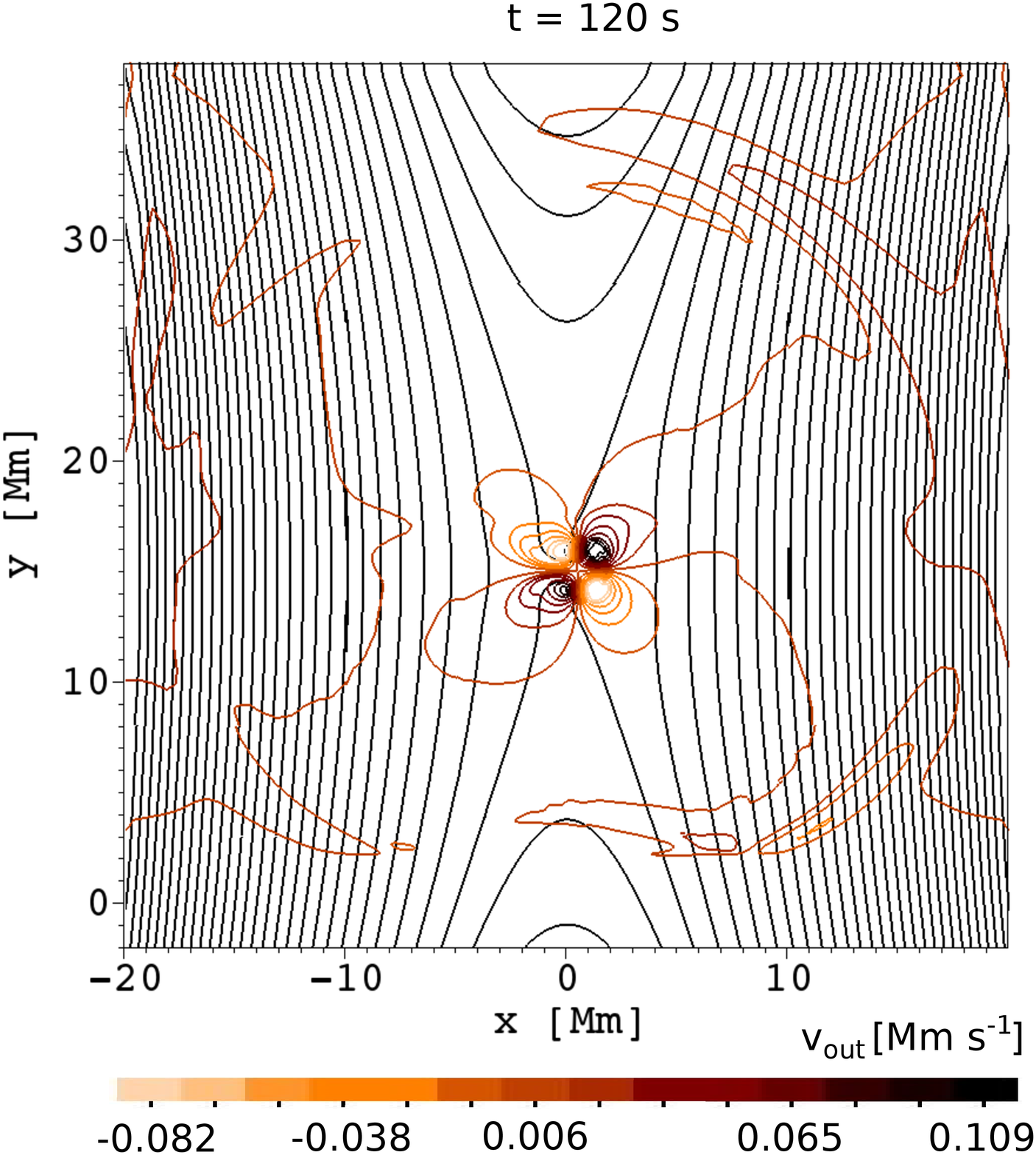}
\hspace{-0.42cm}
\includegraphics[scale=0.6,angle=0,width=4.3cm,height=4.8cm,keepaspectratio]{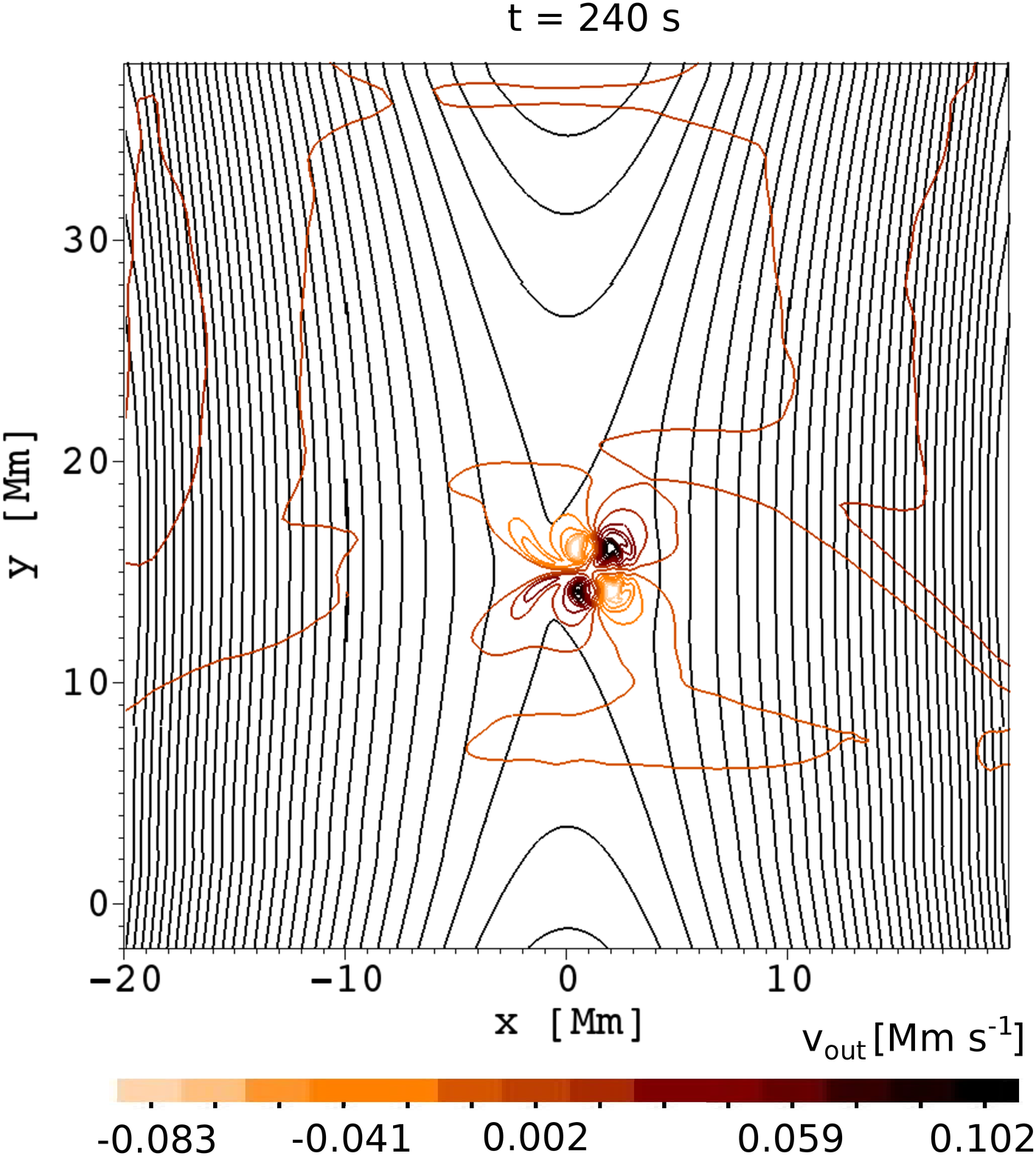}
\hspace{-0.42cm}
\includegraphics[scale=0.6,angle=0,width=4.3cm,height=4.8cm,keepaspectratio]{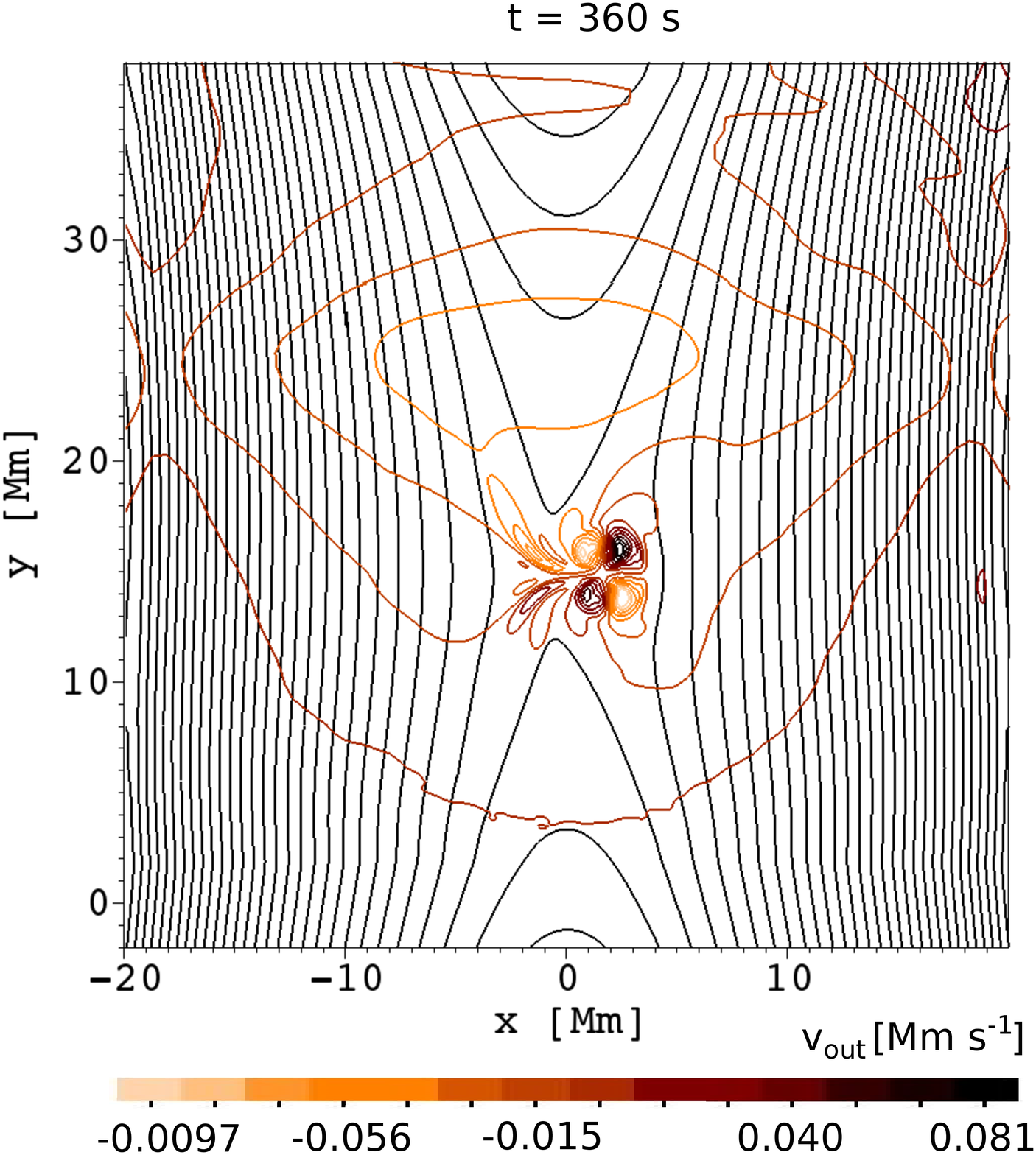}
\hspace{-0.42cm}
\includegraphics[scale=0.6,angle=0,width=4.3cm,height=4.8cm,keepaspectratio]{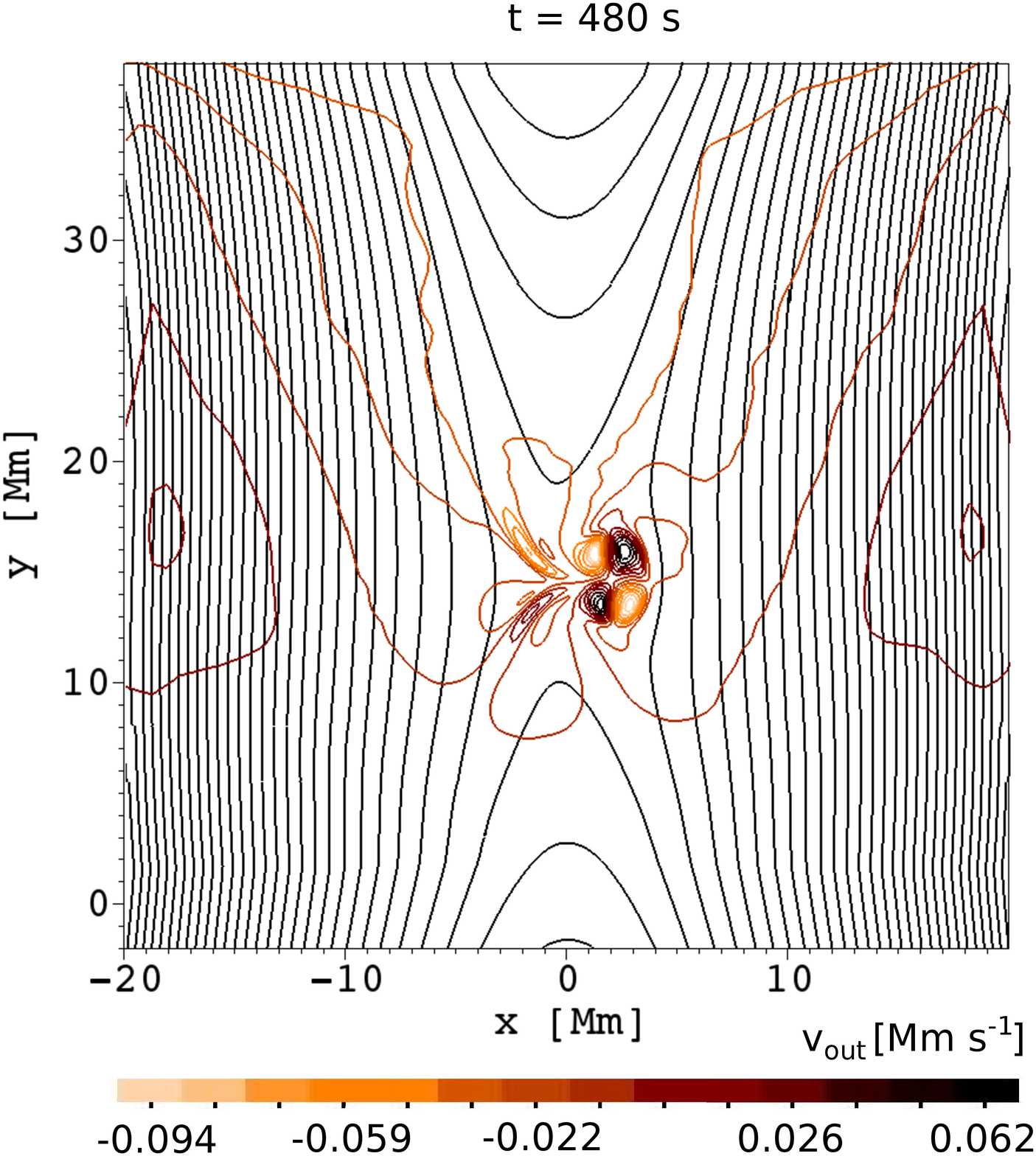}
}
\end{center}
\caption{\small \small Numerical model of the forced magnetic reconnection at X-point. Using FLASH code, the X-point coronal magnetic field configuration has been reproduced on
the observed spatial scales. An external disturbance is imposed on the left side of the X-point, which creates an initial average inflow of $\approx$2.56 km s$^{-1}$ triggering the 
reconnection of the field lines. Thereafter plasma outflows with an average speed of $\approx$12.46 km s$^{-1}$ has been generated during the reconnection in the perpendicular direction of the inflows. The simulated inflows, outflows, and the reconnection rate
match with the observations of the forced reconnection. The various contours with different colours denote the velocities in units of Mm s$^{-1}$. The top and bottom rows display respectively inflow and outflow velocity contours around X-point as overplotted on the reconnection region. The total duration of the reconnection process is $<$10 min in the simulation, which match with the duration of physical processes (inflows, reconnection heating as well as outflows) as established initially in the reconnection region between 13:55 UT and 14:05 UT  (cf., Fig.~4).
The animations $'$fig6-toprow.mp4$'$ and $'$fig6-bottomrow.mp4$'$ respectively display the dynamics of the forced reconnection region with inflow and outflow velocity contours. These animations run for the duration of eight minutes.}
\end{figure*}

The external disturbance propagates across the magnetic field lines as a fast magnetoacoustic perturbation, pushing the magneto-plasma system towards the X-point from the left side (Fig.~6, see $'$fig6-toprow.mp4$'$ and $'$fig6-bottomrow.mp4$'$) with an effective plasma inflows of $\approx$2.52 km s$^{-1}$. This subsequently triggers the forced magnetic reconnection. After the reconnection of the field lines at the X-point, the magneto-plasma system is subjected to the outflows in the perpendicular direction with a speed of $\approx$12.46 km s$^{-1}$. This results in a reconnection rate of $\approx$0.20, which lies close to the observations. This numerical simulation demonstrates that even if a sufficiently appropriate diffusion region is not created in the corona, an external driver can still trigger considerably rapid reconnection (cf., movies 5 \& 6). In the present numerical simulation, we have the value of resistivity, which is sufficient to start physical reconnection, and the numerical effects are almost negligible. The other apsect is that we observe the reconnection very shortly after the start of our simulation confirming that the tiny numerical resistivity does not affect the physical forced reconnection. The reconnection occurs in the present system, where the ratio between the length and width of the current sheet is high.

\begin{figure*}
\begin{center}
\mbox{
\includegraphics[scale=0.6,angle=90,width=8.5cm,height=9.0cm,keepaspectratio]{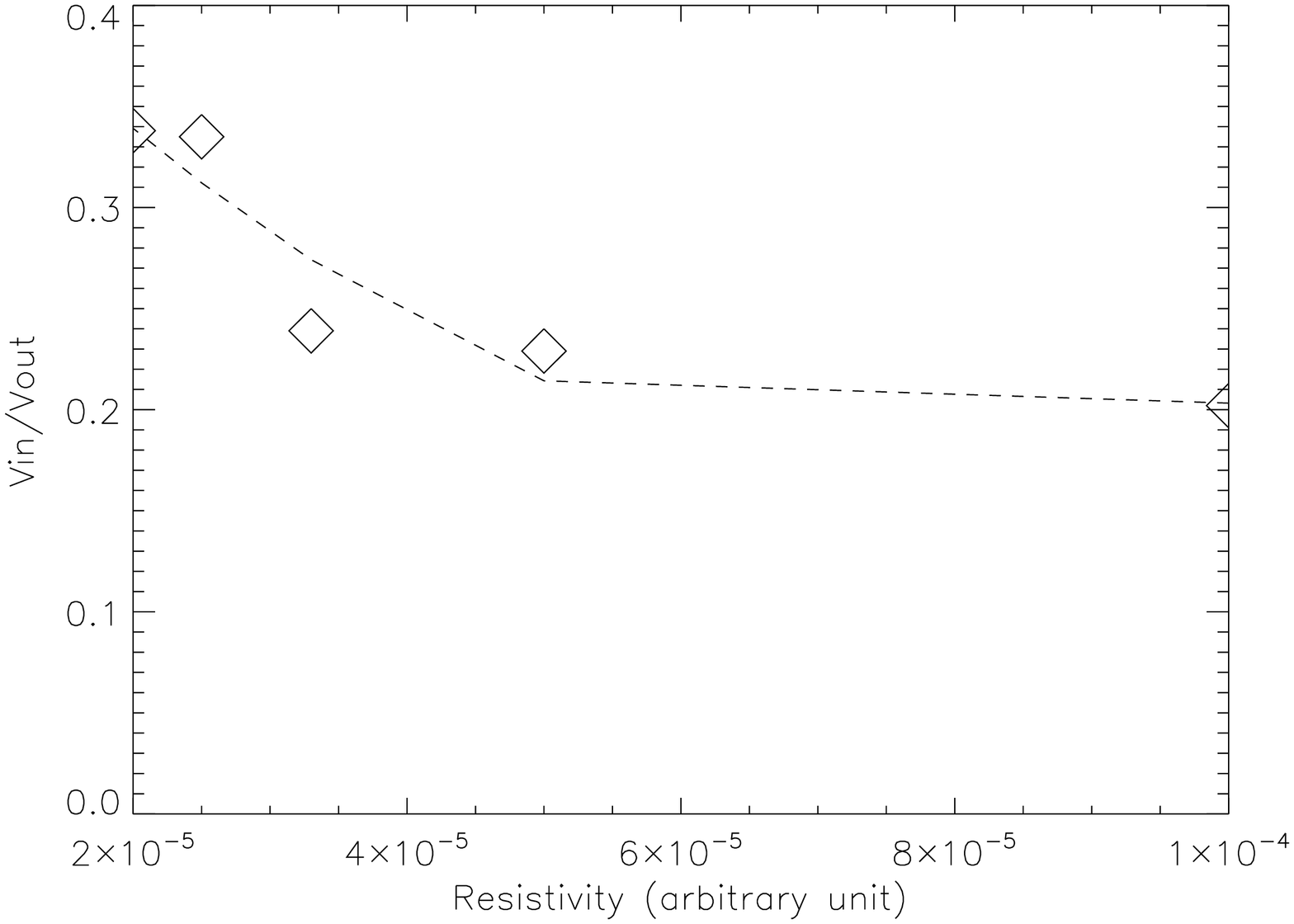}
\includegraphics[scale=0.6,angle=90,width=8.5cm,height=9.0cm,keepaspectratio]{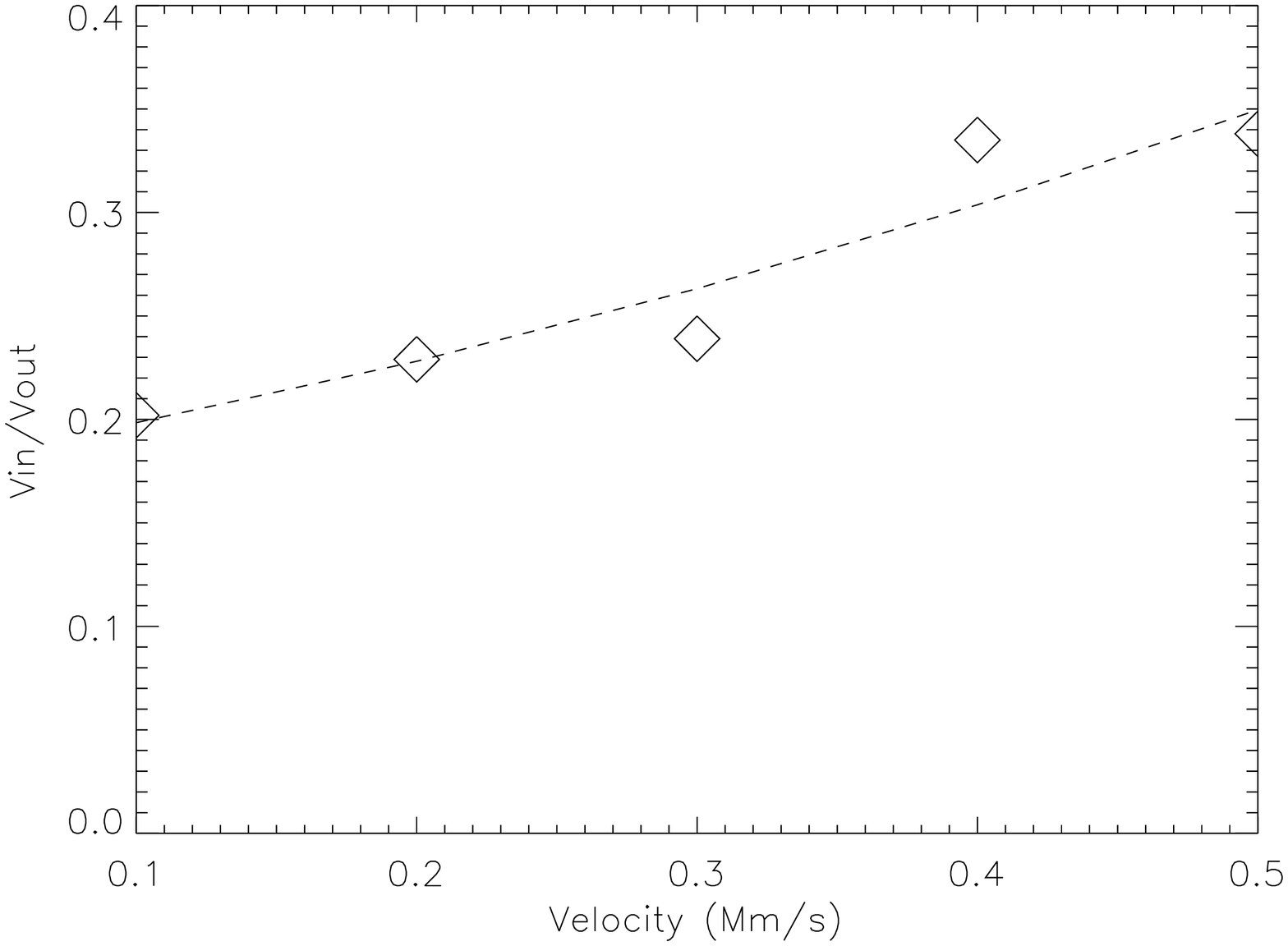}}
\end{center}
\caption{\small \small Left: The magnitude of Vin/Vout with the change in resistivity. Right: The magnitude of Vin/Vout with the change in the strength of the external disturbance, i.e., the initial magnitude of velocity pulse. The third order quadratic fit is also drawn on these two plots.}
\end{figure*}

 To show the physical significance of the observed and modelled forced reconnection in the corona, we performed a parametric study in the numerical simulation on the same spatio-temporal scales as we see in the observations. To understand the role of the external disturbances in the magnetic reconnection region in solar corona, we have performed the parametric study of the numerical simulation. We fix the spatio-temporal scales as depicted in main simulation results and match with the observations. We decrease the resisvity, while increase the magnitude of external velocity pulse mimicking an external disturbance in the vicinity of the X-point (cf., Fig.~7). This indicates that we control more the magnetic reconnection in the localized corona by an external disturbance. Reconnection occurs by producing a pair of plasma inflow and outflow, as well as the formation of a thin current sheet in each case. The estimated reconnection rate increases even after the decrement of the resistivity at the X-point as the external velocity pulse forcing the reconnection increases gradually (cf., Fig.~7). Our main objective of the present work is to study the exclusive observations of the forced reconenction due to an external driver (prominence in our case). We introduce a simple 2-D forced reconnection model of an X-point due to an external driver. We do not specifically implement any solar structure like prominence as an external driver. We wish to understand that even there would be any driver outside how it will exert an impulse and force the resistive X-point into reconnection. We understand the physical significance of this process for various strength of the velocity pulse and resistivity. We find that the reconnection rate (V$_{in}$/V$_{out}$) becomes quite-significant under the influence of an external driver even when we decrease the resistivity of the plasma. Our numerical simulation is a general physical implication of the forced reconnection. Velocity perturbations and magnetic field conditions (initial conditions) of the simulation were chosen from the observations. Therefore, the forced reconnection is established here as a significant mechanism for the rapid release of the magnetic energy in the corona even when natural diffusion does not play an important role.\\
 
Implying what we have mentioned above, Fig.~7 (left-panel) shows that as the resistivity decreases, the reconnection rate Vin/Vout increases. The variations of reconnection rate (V$_{in}$/V$_{out}$) vs resistivity and velocity perturbations are parametric plots based on 5 case studies (Fig.~7). We have compared five case studies of the forced reconnection. The reconnection rate: V$_{in}$/V$_{out}$) in a same magneto-plasma configuration with decreasing resistivity and increasing external velocity disturbances, are compared. The reason is that we decrease resistivity at X-point while simultaneously we increase the magnitude of the external disturbance. Its clear that the implementation of the external driver increases the rate of the reconnection even when the resistivity required for creating normal diffusion region decreases. Fig.~7 (right-panel) shows that as the external disturbance, i.e., initial magnitude of velocity pulse increases, the reconnection rate Vin/Vout also increases. It should be noted that we decrease resistivity at the X-point while simultaneously we increase the magnitude of the external disturbance. Its clear that the implementation of the external driver increases the rate of the reconnection even when the resistivity required for creating normal diffusion region decreases at the X-point.  In conclusion, the forced reconnection in the large-scale solar corona is observationally detected, and its physical significance is established through matching numerical model.

\section{Discussion and Conclusions}

Although the issues related to the magnetic reconnection (e.g., rate, magnitude of diffusivity and resistivity, plasma and magnetic field structuring, amount of stored and released energy) are still highly debated, there is evidence for the unambiguous presence of this physical process which makes it as one of the key mechanisms for coronal heating and plasma eruptions (e.g., Sui et al., 2003;  Lin et al., 2005; Li et al., 2016; Xue et al., 2016). The directly and firstly observed forced reconnection in the present paper does not require the establishment of large magnetic diffusivity in the localized corona as required for the normal reconnection. The specific morphological (a typical length-scale of reconnection) as well as typical magnetic fields also do not influence it significantly. The evolution of the hot plasma is visible in the form of outflowing plasma streaks (Fig.~5c), as well as high temperature emissions (AIA 131 \AA~, Log T$_{e}$=7.0) at the reconnection site (Fig.~5a). However, the bulk energy release in the present case is not as prominent as seen in typical flare sites (Su et al., 2013). This region is a quiescent loop system in the corona where there is no flare related energy stored in magnetic fields. The forced magnetic reconnection does not build a large energy component, however, its stored amount in coronal fields can be liberated over appropriate time-scales typical for the localized heating and transient processes (Shibata \& Magara, 2011; Vekstein, 2017). We have also introduced a forced reconnection model to support its exclusive new observations presented in our paper. We achieved out scientific objectives with the present description of the model, which also emphasize the parametric study on the estimation of the reconnection rate w.r.t. resistivity as well as strength of the external disturbances. It draws an important new conclusion that external forcing of the magnetic reconnection may play an important role in determining the reconnection rate even in the case when resitivity is comparatively small.

The highly dynamic and complex solar corona can be inevitably subjected to such forced reconnection at diverse spatio-temporal scales when external disturbances act on the partially or fully established reconnection regions, thus making it a significant physical mechanism for a variety of dynamical plasma processes in the solar corona (Jain et al., 2005; Potter et al., 2019). These first observational clues to the forced reconnection can also be extended to the laboratory plasma to constrain the behaviour of highly diffusive plasmas (Yamada et al., 2014). The forced magnetic reconnection responsible for the formation of current sheet in the MHD stable configuration. The chromospheric/prominence system is dominated by cool, partially ionized and collision dominated plasma. Therefore, the most of the energy release due to magnetic reconnection may be consumed by these plasma if they are present in the vicinity of energy release site (e.g., Chen et al. 2001.; Chen \& Ding 2006; Jess et al. 2010). The energy release by the forced magnetic reconnection is useful for surrounding plasma heating (e.g., Vekstein \& Jain 2005; Vekstein 2017). Xue et al. (2018) have observed a current sheet in small scale magnetic reconnection event. They found that the average temperature of the current sheet region is lower than 1.6 MK. The lower temperature, reconnection in partially ionized plasma may be different from more commonly studied coronal reconnection. In the present paper, we observed the prominence driven forced magnetic reconnection. Therefore, we may conclude that the prominence (partially ionized, cool and dense region) consumed the energy generated by the forced magnetic reconnection, which contain an elongated  current sheet (2.5 MK).

\section{Acknowledgment}
Authors thank the referee for his/her significant remarks that impproved the masnucript considerably. AKS acknowledges UKIERI project grant, and the Advanced Solar Computational \& Analyses Laboratory (ASCAL). P. J. acknowledges support from Grant 16-13277S of the Grant Agency of the
Czech Republic. AKS, DB, and BND acknowledge the Indo-US (IUSSTF/JC-2016/011) for the support in their research. TS and HT are supported by NSFC grants 41574166 and 11790304 (11790300). Armagh Observatory and Planetarium is grant-aided by the N. Ireland Department of Communities. JGD acknowledges the DJEI/DES/SFI/HEA Irish Centre for High-End Computing (ICHEC) for the provision of computing facilities and support. JGD also thanks STFC for PATT T\&S and the SOLARNET project which is supported by the European Commission's FP7 Capacities Programme
under Grant Agreement number 312495 for T\&S.


\begin{thebibliography}{99}
\bibitem[Baty(2000)]{2000A&A...353.1074B} Baty, H.\ 2000, \aap, 353, 1074 
\bibitem[Beidler et al.(2017)]{2017PhPl...24e2508B} Beidler, M.~T., Callen, J.~D., Hegna, C.~C., \& Sovinec, C.~R.\ 2017, Physics of Plasmas, 24, 052508
\bibitem[Birn et al.(2005)]{2005GeoRL..32.6105B} Birn, J., Galsgaard, K., Hesse, M., et al.\ 2005, \grl, 32, L06105 
\bibitem[Cargill \& Klimchuk(2004)]{2004ApJ...605..911C} Cargill, P.~J., \& Klimchuk, J.~A.\ 2004, \apj, 605, 911
\bibitem[Chen et al.(2001)]{2001ChJAA...1..176C} Chen, P.-F., Fang, C., \& Ding, M.-D.~D.\ 2001, \cjaa, 1, 176 
\bibitem[Chung(2002)]{2002cfd..book.....C} Chung, T.~J.\ 2002, Computational Fluid Dynamics, by T.~J.~Chung, pp.~1036.~ISBN 0521594162.~Cambridge, UK: Cambridge University Press, March 2002., 10
\bibitem[Chen \& Ding(2006)]{2006ApJ...641.1217C} Chen, Q.~R., \& Ding, M.~D.\ 2006, \apj, 641, 1217
\bibitem[De Pontieu et al.(2007)]{2007Sci...318.1574D} De Pontieu, B.~W., Carlsson, M., et al.\ 2007, Science, 318, 1574 
\bibitem[Dewar et al.(2013)]{2013PhPl...20h2103D} Dewar, R.~L., Bhattacharjee, A., Kulsrud, R.~M., \& Wright, A.~M.\ 2013, Physics of Plasmas, 20, 082103
 \bibitem[Fitzpatrick(2003)]{2003PhPl...10.2304F} Fitzpatrick, R.\ 2003, Physics of Plasmas, 10, 2304 
\bibitem[Fryxell et al.(2000)]{2000ApJS..131..273F} Fryxell, B., Olson, K., Ricker, P., et al.\ 2000, \apjs, 131, 273
\bibitem[Hahm \& Kulsrud (1985)]{1985PhPl...28.2412} Hahm, T. S. \& Kulsrud, R. M., 1985, Physics of Plasma, 28, 2412
\bibitem[Hannah \& Kontar(2012)]{2012A&A...539A.146H} Hannah, I.~G., \& Kontar, E.~P.\ 2012, \aap, 539, A146
\bibitem[Jain et al.(2005)]{2005PhPl...12a2904J} Jain, R., Browning, P., \& Kusano, K.\ 2005, Physics of Plasmas, 12, 012904
\bibitem[Jel{\'{\i}}nek et al.(2015)]{2015ApJ...812..105J} Jel{\'{\i}}nek, P., Karlick{\'y}, M., \& Murawski, K.\ 2015, \apj, 812, 105
\bibitem[Jel{\'{\i}}nek et al.(2017)]{2017ApJ...847...98J} Jel{\'{\i}}nek, P., Karlick{\'y}, M., Van Doorsselaere, T., \& B{\'a}rta, M.\ 2017, \apj, 847, 98   
\bibitem[Jess et al.(2009)]{2009Sci...323.1582J} Jess, D.~B., Mathioudakis, M., Erd{\'e}lyi, R., et al.\ 2009, Science, 323, 1582
\bibitem[Jess et al.(2010)]{2010ApJ...712L.111J} Jess, D.~B., Mathioudakis, M., Browning, P.~K., Crockett, P.~J., \& Keenan, F.~P.\ 2010, \apjl, 712, L111
\bibitem[Klimchuk(2015)]{2015RSPTA.37340256K} Klimchuk, J.~A.\ 2015, Philosophical Transactions of the Royal Society of London Series A, 373, 20140256
\bibitem[Lemen et al.(2012)]{2012SoPh..275...17L} Lemen, J.~R., Title, A.~M., Akin, D.~J., et al.\ 2012, \solphys, 275, 17
\bibitem[Li et al.(2016)]{2016NatPh..12..847L} Li, L., Zhang, J., Peter, H., et al.\ 2016, Nature Physics, 12, 847 
\bibitem[Lin et al.(2005)]{2005ApJ...622.1251L} Lin, J., Ko, Y.-K., Sui, L., et al.\ 2005, \apj, 622, 1251
\bibitem[Lee(2013)]{2013JCoPh.243..269L} Lee, D.\ 2013, Journal of Computational Physics, 243, 269
\bibitem[Litvinenko(1999)]{1999ApJ...515..435L} Litvinenko, Y.~E.\ 1999, \apj, 515, 435 
\bibitem[Litvinenko et al.(2007)]{2007ApJ...662.1302L} Litvinenko, Y.~E., Chae, J., \& Park, S.-Y.\ 2007, \apj, 662, 1302
\bibitem[Mart{\'{\i}}nez-Sykora et al.(2017)]{2017Sci...356.1269M} Mart{\'{\i}}nez-Sykora, J., De Pontieu, B., Hansteen, V.~H., et al.\ 2017, Science, 356, 1269
\bibitem[McIntosh et al.(2011)]{2011Natur.475..477M} McIntosh, S.~W., de Pontieu, B., Carlsson, M., et al.\ 2011, \nat, 475, 477
\bibitem[Parnell et al.(1997)]{1997GApFD..84..245P} Parnell, C.~E., Neukirch, T., Smith, J.~M., \& Priest, E.~R.\ 1997, Geophysical and Astrophysical Fluid Dynamics, 84, 245 
\bibitem[Potter et al.(2019)]{2019arXiv190102392P} Potter, M., Browning, P., \& Gordovskyy, M.\ 2019, arXiv:1901.02392 
\bibitem[Priest \& Forbes(2007)]{2007mare.book.....P} Priest, E., \& Forbes, T.\ 2007, Magnetic Reconnection, by Eric Priest , Terry Forbes, Cambridge, UK: Cambridge University Press, 2007
\bibitem[Priest(2014)]{2014masu.book.....P} Priest, E.\ 2014, Magnetohydrodynamics of the Sun, by Eric Priest, Cambridge, UK: Cambridge University Press, 2014

\bibitem[Sakai et al.(1984)]{1984SoPh...91..103S} Sakai, J., Tajima, T., \& Brunel, F.\ 1984, \solphys, 91, 103 
\bibitem[Savage et al.(2012)]{2012ApJ...754...13S} Savage, S.~L., Holman, G., Reeves, K.~K., et al.\ 2012, \apj, 754, 13 
\bibitem[Schwenn(2006)]{2006LRSP....3....2S} Schwenn, R.\ 2006, Living Reviews in Solar Physics, 3, 2 
\bibitem[Solov'ev(2010)]{2010ARep...54...86S} Solov'ev, A.~A.\ 2010, Astronomy Reports, 54, 86 
\bibitem[Shibata \& Magara(2011)]{2011LRSP....8....6S} Shibata, K., \& Magara, T.\ 2011, Living Reviews in Solar Physics, 8, 6
\bibitem[Srivastava et al.(2017)]{2017NatSR...743147S} Srivastava, A.~K., Shetye, J., Murawski, K., et al.\ 2017, Nat. Sci. Rep., 7, 43147
\bibitem[Srivastava et al.(2018)]{2018NatAs...2..951S} Srivastava, A.~K., Murawski, K., Ku{\'z}ma, B., et al.\ 2018, Nature Astronomy, 2, 951
\bibitem[Su et al.(2013)]{2013NatPh...9..489S} Su, Y., Veronig, A.~M., Holman, G.~D., et al.\ 2013, Nature Physics, 9, 489
\bibitem[Sui \& Holman(2003)]{2003ApJ...596L.251S} Sui, L., \& Holman, G.~D.\ 2003, \apjl, 596, L251 
\bibitem[Toro(2006)]{2006ARA&A..15..363W} Toro, E.F.\ 2006, Int. J. Num. Meth. Fluids, 52, 433
\bibitem[Takasao et al.(2012)]{2012ApJ...745L...6T} Takasao, S., Asai, A., Isobe, H., \& Shibata, K.\ 2012, \apjl, 745, L6 
\bibitem[Vernazza et al.(1981)]{1981ApJS...45..635V} Vernazza, J.~E., Avrett, E.~H., \& Loeser, R.\ 1981, \apjs, 45, 635 
 \bibitem[Vekstein \& Jain(1998)]{1998PhPl....5.1506V} Vekstein, G.~E., \& Jain, R.\ 1998, Physics of Plasmas, 5, 1506 
\bibitem[Vekstein(2017)]{2017JPlPh..83e2001V} Vekstein, G.\ 2017, Journal of Plasma Physics, 83, 205830501
\bibitem[Wedemeyer-B{\"o}hm et al.(2012)]{2012Natur.486..505W} Wedemeyer-B{\"o}hm, S., Scullion, E., Steiner, O., et al.\ 2012, \nat, 486, 505
\bibitem[Withbroe \& Noyes(1977)]{1977ARA&A..15..363W} Withbroe, G.~L., \& Noyes, R.~W.\ 1977, \araa, 15, 363
\bibitem[White(1984)]{1984bpp..conf..611W} White, R.~B.\ 1984, Basic Plasma Physics: Selected Chapters, Handbook of Plasma Physics, Volume 1, 611 
\bibitem[Xue et al.(2016)]{2016NatCo...711837X} Xue, Z., Yan, X., Cheng, X., et al.\ 2016, Nature Communications, 7, 11837 
\bibitem[Xue et al.(2018)]{2018ApJ...858L...4X} Xue, Z., Yan, X., Yang, L., et al.\ 2018, \apjl, 858, L4 
\bibitem[Yamada et al.(2010)]{2010RvMP...82..603Y} Yamada, M., Kulsrud, R., \& Ji, H.\ 2010, Reviews of Modern Physics, 82, 603 
\bibitem[Yamada et al.(2014)]{201tNatCom...5..4774Y} Yamada, M., Yoo, J., Jara-Almonte, J., Ji, H., Kulsrud, R.M., \& Myers, C.E.\ 2014, Nat. Comm.,5, 4774 
\bibitem[Yan et al.(2018)]{2018ApJ...853L..18Y} Yan, X.~L., Yang, L.~H., Xue, Z.~K., et al.\ 2018, \apjl, 853, L18 
\end{thebibliography}
\end{document}